\documentclass[aps,showpacs,twocolumn,floatfix,amsmath,amssymb,nofootinbib]{revtex4-1}
\usepackage{amssymb}
\usepackage{amsmath}
\usepackage{epsfig}
\usepackage{graphicx}
\usepackage{latexsym}
\usepackage{bm,latexsym}
\usepackage{mathrsfs}
\usepackage{float}
\usepackage{pbox}

\def\Journal#1#2#3#4{{#1} {\bf #2}, #3 (#4)}
\def\AA{{Astron. Astrophys.}}

\def\AJ{{Astron. J.}}

\def\APJ{{Astrophys. J.}}

\def\CQG{{Class. Quantum Grav.}}

\def\GRG{{Gen. Relativ. Gravit.}}
\def\IJMPD{{Int. Jour. Mod. Phys. D}}
\def\JCAP{{JCAP}}

\def\LRR{{Living Rev. Rel.}}

\def\PRL{Phys. Rev. Lett.}

\def\PRD{{\em Phys. Rev.} D}

\def\RMP{{Rev. Mod. Phys.}}

\begin{document}

\title{Cosmic acceleration in asymptotically Ricci flat Universe}

\author{Luisa G. Jaime}
\email{luisa@nucleares.unam.mx}

\author{Marcelo Salgado}
\email{marcelo@nucleares.unam.mx}

\affiliation{Instituto de Ciencias Nucleares, Universidad Nacional Aut\'onoma de M\'exico.}


\date{\today}

    
\begin{abstract}
{We analyze the evolution of a Friedmann-Robertson-Walker spacetime within the framework of $f(R)$ metric gravity using an exponential model. We show that $f(R)$ gravity may lead to a vanishing effective cosmological constant in the far future ({\it i.e.} $R\rightarrow 0$) and yet produce a transient accelerated expansion at present time with a potentially viable cosmological history. This is in contrast with 
several $f(R)$ models which, while viable, produce in general a non-vanishing effective cosmological constant asymptotically in time
($R\rightarrow 4\Lambda_{\rm eff}$). We also show that relativistic {\it stars} in asymptotically flat spacetimes can be supported within this framework without encountering any singularity, notably in the Ricci scalar $R$.}
\end{abstract}


\pacs{
04.50.Kd, 
95.36.+x  
04.40.Dg, 
}


\maketitle


\section{Introduction}
\label{sec:introduction}

$f(R)$ gravity has been proposed recently as a {\it natural} mechanism to generate an effective cosmological constant even if $f(0) \equiv 0$ 
(for a review see~\cite{Sotiriou2010,Capozziello2008a,deFelice2010,Jaime2012a} and references therein) and also as a model for cosmic 
inflation in the early Universe~\cite{Starobinsky1979}. This effective cosmological constant is then capable to explain the current accelerated expansion of the Universe~\cite{Perlmutter1999, Riess1998, Amanullah2010}. The heuristic argument that allows to appreciate this property in simple grounds is as follows: the field equations of this kind of theories give rise to an evolution equation for the Ricci scalar $R$ [cf. Eq.~(\ref{traceR}) of Section~\ref{sec:f(R)}]. If the trace $T$ of the energy momentum of matter vanishes, then this equation admits $R=R_1= const$ as solution when $R_1$ is a solution of the algebraic 
equation $2f(R_1)-R_1f_R(R_1)=0$ (provided $f_{RR}(R_1)\neq 0$) where the subindex $R$ refers to a derivative of $f(R)$ with respect to such variable. When this solution is 
replaced in the field equations, the latter become the Einstein field equations endowed with an effective cosmological constant $\Lambda_{\rm eff}= R_1/4$ and an effective gravitational constant~\cite{Jaime2016}. Now, even though $T$ does not vanish in general, notably, during the matter dominated epoch, detailed numerical analyses of the Friedmann-Robertson-Walker (FRW) spacetime within $f(R)$ gravity~\cite{Jaime2012a} show that $R$ evolves from a given value in the past (say from a matter dominated Universe where $T\approx -\rho_{\rm matt}$) to the attractor solution  $R_1$ in the future while $\rho_{\rm matt}\sim a^{-3}$ vanishes as the scale factor grows. Therefore, asymptotically in time $R\rightarrow R_1$ and $T\rightarrow 0$. Thus, $f(R)$ gravity generates in a dynamical fashion an effective cosmological constant in the future. Of course at present time $R\approx R_1$ and $T\approx -0.3$ (in units of critical energy density), so in fact the effective equation of state of the {\it geometric dark energy} $\omega_X$ mimicked by $f(R)$ gravity is not constant but evolves in cosmic time such that $\omega_X\rightarrow -1$ as $t\rightarrow \infty$, but $\omega_X\approx -1$ at present time~\cite{Jaime2014}. Moreover, thanks to this behavior, several $f(R)$ models can be in good agreement with the luminosity distance-redshift relation inferred from type I supernovae (SNIa)~\cite{Jaime2012a}.

Now, as regards the Solar System constraints within $f(R)$ gravity, this is a particularly subtle issue which has been the object of a long debate in the past 
but that seems to be more or less settle today for certain models, but not entirely in general (cf.~\cite{Faulkner2007}). 
The point is that $f(R)$ gravity can be recasted as a kind of Brans-Dicke (BD) theory with a parameter $\omega_{\rm BD}\equiv 0$. This is because the scalar-degree of freedom has a vanishing kinetic contribution. Since observations require $\omega_{\rm BD} \gtrsim 4\times 10^4$, the naive conclusion is that $f(R)$ gravity is blatantly ruled out. The caveat of this argument is that in fact the emergent scalar-tensor theory is {\it not} exactly the original BD theory 
with $\omega_{\rm BD}\equiv 0$ but it is endowed with a scalar-field potential. Therefore, if the potential has certain features, the theory can exhibit a {\it chameleon} like behavior~\cite{Khoury2004} which is responsible for suppressing the large deviations from general relativity (GR) in regions around the Sun. As a consequence, some $f(R)$ models can survive the Solar System tests~\cite{Hu2007,Faulkner2007}. However, in order to check that this indeed happens require, in principle, a very detailed numerical analysis that involves a high numerical accuracy, and which has to be done in a case-by-case basis, i.e., for each specific $f(R)$ model.

A similar kind of accuracy is involved when constructing realistic neutron stars. This can be seen from the fact that $f(R)$ models built to explain the cosmic acceleration involves a natural length scale $\ell\sim 1/\sqrt{\Lambda_{\rm eff}}$. This scale is huge compared to the length scales involved in neutron stars, which are of the order of ten kilometers. These lengths scales can be translated into density scales, which are of the order $\Lambda_{\rm eff}/G_0\sim\rho_{\rm crit}$, while the energy-densities characteristic of a neutron star are several orders of magnitude larger than the cosmological values during most of the cosmic evolution. Therefore, handling such contrasts of densities are numerically challenging.

In the case of GR endowed with a cosmological constant, usually one constructs neutron star models embedded in a spacetime that is 
asymptotically flat (AF) by simply neglecting the cosmological constant. However, in $f(R)$ gravity one cannot simply set $\Lambda_{\rm eff}\equiv 0$ as this quantity emerges dynamically. Moreover, most of alternative $f(R)$ models have an intrinsic scale $R_*\sim \Lambda_{\rm eff}$ which cannot be set to zero (see Section~\ref{sec:f(R) model}).

A partial solution to this technical problem consists in trying to construct ``compact'' objects that are large compared to neutron stars so as to avoid 
the handling of two different scales, but which are however, relativistic in the sense that its pressure is large and comparable to its energy density and whose mass-to-radius ratio $G_0 M/c^2 R$ is similar to the one of a neutron star. 

These relativistic objects can be used as a testbed for $f(R)$ theory in dealing with the {\it strong} gravity regime. 
It turns, however, that even in this simplified scenario some attempts to find relativistic objects failed due to the appearance of a {\it curvature singularity}~\cite{Kobayashi2008,Kobayashi2009,Babichev2009,Upadhye2009}. Furthermore, detailed analyses showed that 
such a drawback can be in fact avoided~\cite{Jaime2011}. However, this entails changing the original parameters of the $f(R)$ model, which can put in jeopardy the cosmological and even the Solar System tests. In other words, as of today, there is no single $f(R)$ self-consistent model compatible with all the tests of general relativity while satisfying the condition $f(0)=0$, i.e., a $f(R)$ model without the inclusion of an explicit cosmological constant\footnote{Several authors have used $f(R)$ gravity to test its implications in the strong gravity regime but without invoking the theory as a model for dark-energy~\cite{neutronstarsf(R)}
. Thus, the intrinsic scale $R_*$ involved in such models has no relationship with the cosmological scales and the technical problems alluded in the main text are avoided.}. Otherwise, such problems can be avoided by simply selecting $f_{\rm GR}(R)= R-2\Lambda$, which corresponds to GR with a cosmological constant.

At this point one can then argue, why not selecting $f_{\rm GR}(R)$ , which is the simplest and the most successful model. Perhaps the most honest answer we can give {\it a posteriori} is that there are some measurements of the Hubble expansion $H$ at different epochs that are in {\it mild} tension with the $\Lambda$CDM model~\cite{boss2014}. Moreover in~\cite{Zhao2017} a statistical analysis shows that these tensions might be relieved with a dynamical dark energy and in~\cite{Jaime2015} it was argued that $f(R)$ gravity could help to alleviate them. So, if such dark energy turns out to be varying in cosmic time with an equation of state (EOS) 
different from the value $\omega_{\rm DE}=-1$, then we would have a concrete prediction other than the standard $\Lambda$CDM model. Therefore, modified gravity has the potential of dealing with such tensions in a very well defined manner. It is then worth pursuing the analysis of such possibility, even if at the end the $\Lambda$CDM model is confirmed by future experiments and the tensions are solved by a better statistics.

In this article we want to explore an exponential $f(R)$ model such that $R_1\equiv 0$ and thus, $\Lambda_{\rm eff}\equiv 0$. That is, a model where the attractor solution in cosmology is the one where the effective cosmological constant vanishes asymptotically in time, but where the transient behavior of $R$ is such that its value today is close to the observed value $R\sim 4\Lambda$ as predicted by the $\Lambda$CDM model. Furthermore, in this kind of $f(R)$ model neutron stars can be embedded naturally in an AF spacetime where $R\rightarrow 0$ at spatial infinity. 

It is important to emphasize that the exponential $f(R)$ model that we use contrasts with seemingly related models that are, however, cosmologically nonviable, like the popular $R^n$ model ~\cite{Amendola2007a,Amendola2007b,Amendola2007c,Jaime2013}. Whereas most of the cosmologically viable models analyzed so far posses a non zero $\Lambda_{\rm eff}$ ~\cite{Starobinsky2007,Hu2007,Miranda2009,Jaime2012a}. In fact, some of these $f(R)$ models admit also $R_1=0$ as a solution when $T=0$, 
but this is {\it not} an attractor solution in cosmology. Besides, even if one tried (somehow) to reinforce the asymptotic solution 
$R_1=0$ in such models~\cite{Starobinsky2007,Hu2007}, one would encounter a singularity in the equation of motion for $R$ at the place where $f_{RR}$ vanishes 
[cf. Eq.~(\ref{traceR}) of Section~\ref{sec:f(R)}] before reaching the value $R=R_1=0$.

The exponential model that we analyze has $R_1=0$ as an attractor solution in cosmology (Section~\ref{sec:cosmologysols}), and in addition $f_{RR}$ is 
positive definite. These features make also possible to construct AF spacetimes, namely, solutions of compact objects with this kind 
of asymptotics (Section~\ref{sec:SSSsols}).


\section{$f(R)$ gravity}
\label{sec:f(R)}
The field equation in $f(R)$ theory is derived from the following action:
\begin{equation}
\label{f(R)}
S[g_{ab},{\mbox{\boldmath{$\psi$}}}] =
\!\! \int \!\! \frac{f(R)}{2\kappa} \sqrt{-g} \: d^4 x 
+ S_{\rm matt}[g_{ab}, {\mbox{\boldmath{$\psi$}}}] \; ,
\end{equation}
where  $\kappa \equiv 8\pi G_0$ ($c=1$), $f(R)$ is an {\it a priori} arbitrary function of the Ricci scalar $R$, and ${\mbox{\boldmath{$\psi$}}}$ 
represents schematically the matter fields.

The field equation arising from variation of the action~(\ref{f(R)}) with respect to the metric is
\begin{equation}
\label{fieldeq1}
f_R R_{ab} -\frac{1}{2}fg_{ab} - 
\left(\nabla_a \nabla_b - g_{ab}\Box\right)f_R= \kappa T_{ab}\,\,,
\end{equation}
where $f_R=\partial_R f$, $\Box= g^{ab}\nabla_a\nabla_b$ is the covariant D'Alambertian and $T_{ab}$ is the energy-momentum 
tensor of matter. From this equation it is not difficult to show that $T_{ab}$ is conserved, i.e., $\nabla^a T_{ab}=0$~\cite{Jaime2016}. 
We rewrite Eq.~(\ref{fieldeq1}) as
\begin{eqnarray}
\label{fieldeq2}
&& f_R G_{ab} - f_{RR} \nabla_a \nabla_b R - 
 f_{RRR} (\nabla_aR)(\nabla_b R) \nonumber \\
&+&  g_{ab}\left[\frac{1}{2}\left(Rf_R- f\right)
+ f_{RR} \Box R + f_{RRR} (\nabla R)^2\right] \nonumber\\
&    & = \kappa T_{ab}\,\,,
\end{eqnarray}
where $G_{ab}= R_{ab}-g_{ab}R/2$ is the Einstein tensor and $(\nabla R)^2:= g^{ab}(\nabla_aR)(\nabla_b R)$. 
The trace of equation Eq.~(\ref{fieldeq2}) yields
\begin{equation}
\label{traceR}
\Box R= \frac{1}{3 f_{RR}}\left[\rule{0mm}{0.4cm}\kappa T - 3 f_{RRR} (\nabla R)^2 + 2f- Rf_R \right]\,\,\,,
\end{equation}
where $T:= T^a_{\,\,a}$. Using~(\ref{traceR}) in~(\ref{fieldeq2}) we find

\begin{eqnarray}
\label{fieldeq3}
G_{ab} &=& \frac{1}{f_R}\Bigl{[} f_{RR} \nabla_a \nabla_b R +
 f_{RRR} (\nabla_aR)(\nabla_b R) \nonumber \\
 &       &- \frac{g_{ab}}{6}\Big{(} Rf_R+ f + 2\kappa T \Big{)} 
+ \kappa T_{ab} \Bigl{]} \; .
\end{eqnarray}

We use Eqs.~(\ref{traceR}) and ~(\ref{fieldeq3}) as the fundamental field equations in this paper, much along the lines described in~\cite{Jaime2011,Jaime2012a}.
 
As stressed before, we see that Eq.~(\ref{traceR}) admits $R=R_1= const$ as a particular solution when the energy-momentum tensor of matter is 
traceless ($T\equiv 0$) provided $R_1$ is an algebraic root of the function:
\begin{equation}
 d{{\cal V}}/dR:= (2f- Rf_R)/(3f_{RR}) \; .
\end{equation}

Aside from some ``exceptional'' cases where both the numerator $2f- Rf_R$ and the denominator $f_{RR}$ vanish at $R_1$ (for example $R^n$ model~\cite{Jaime2013}), 
in general, if $f_{RR}(R_1)\neq 0$, $R_1$ is only a root of:
\begin{equation}
\label{dV}
 dV/dR:= (2f- Rf_R)/3 \; .
\end{equation}

In this instance, the ``potential'' $V(R)= -R f(R)/3 + \int^R f(x) dx$  is useful to track the critical points at $R_1$, notably, the extrema (maxima or minima). So, the three possibilities is $R_1$ to be positive, negative or zero, which are associated with a de Sitter, anti de Sitter or Ricci flat, ``points'', respectively. Clearly the exact location of the critical points depends on the form of the $f(R)$ model and also on the specific value of the parameters involved in this function. 

In the following Section we describe the specific exponential  $f(R)$ model used in this work, and then test it in a  
cosmological scenario and within the context of relativistic objects in hydrostatic equilibrium in order to asses some of 
its most basic viability.

\section{The exponential $f(R)$ model}
\label{sec:f(R) model}

\begin{figure}
\begin{center}
\includegraphics[scale=0.7]{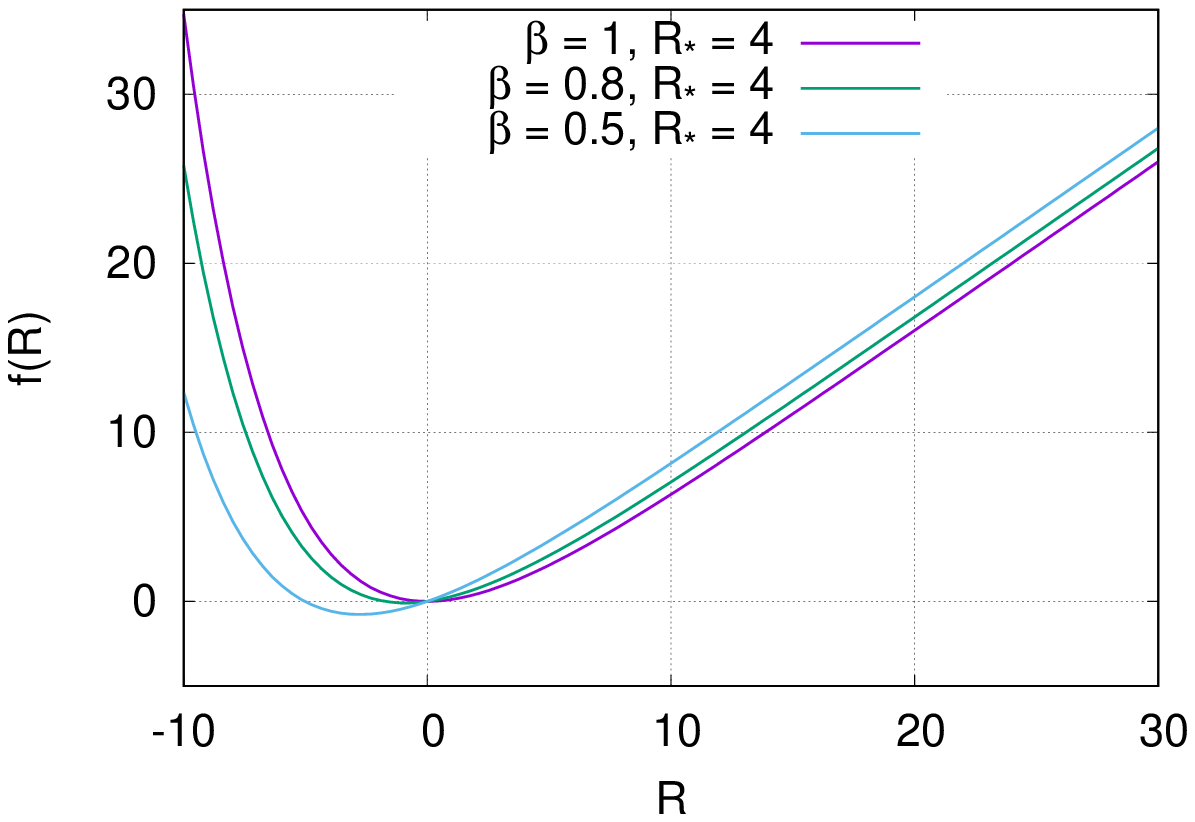}
\includegraphics[scale=0.7]{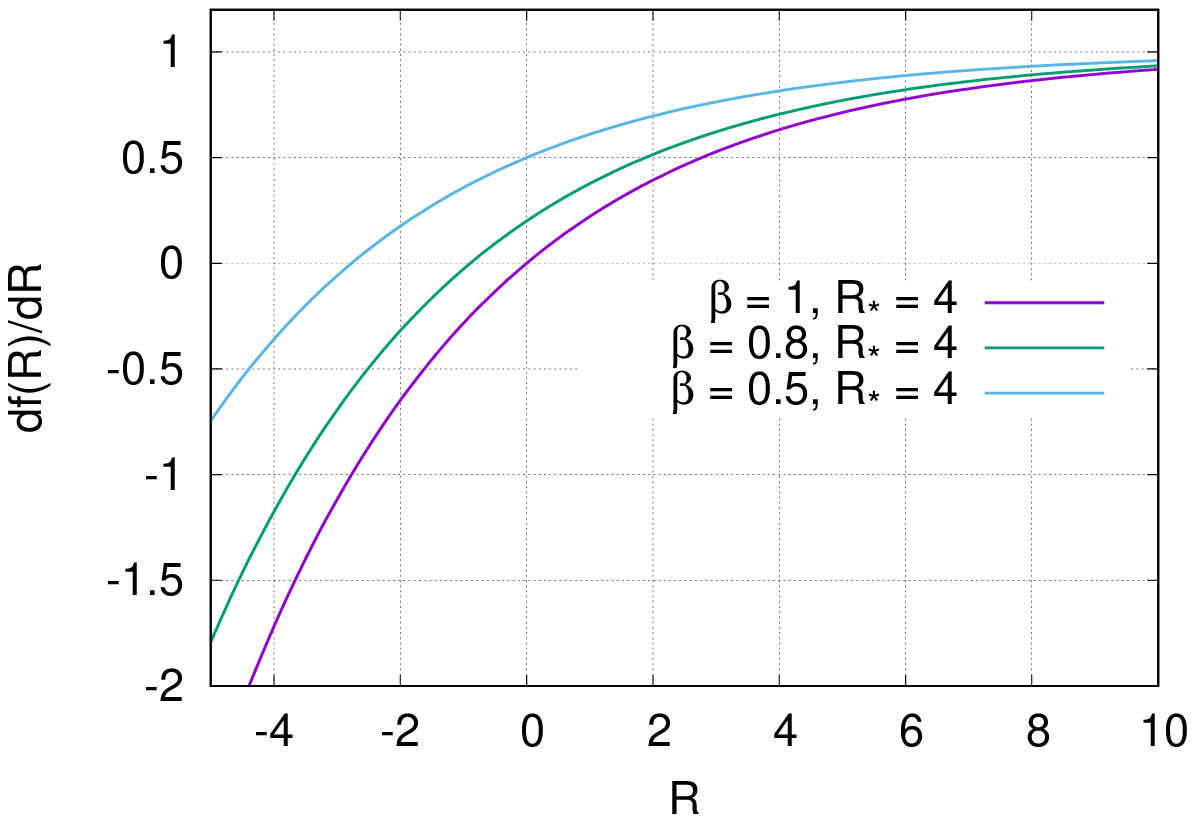}
\caption{Top: The exponential $f(R)$ model for $\beta = 1.0,\, 0.8\,$ and $0.5$ with $R_*=4H_{0}^{2}$ ($R$ and $f(R)$ in units of $H_0^2$). Bottom: First derivative of $f(R)$ for the same parameters.}
\label{fig:f-f1}
\end{center}
\end{figure}

As mentioned in the Introduction, several $f(R)$ models have been proposed in the past in order to produce an accelerated expansion in the Universe with a non vanishing effective cosmological constant. However, in this work we focus on a specific model that allows for asymptotically Ricci flat solutions 
without encountering any singularity. We thus assume the model:
\begin{equation}
\label{ecu-fR-exp}
f(R)= R - \beta R_{*} (1- e^{-R/R_{*}})
\end{equation}
where $R_*$ and $\beta$ are positive parameters. In particular, $R_*$ fixes the scale, and we take $R_*= 4 H_0^2$, while $\beta$ is dimensionless. 
Here $H_0$ stands for the current Hubble expansion. The value of $\beta$ determines the existence of several critical points for the ``potential'' $V(R)$ at $R_1$. In particular for $0<\beta\leq 1$, the potential $V(R)$ has one minimum at $R_1=0$ where in addition $f(R_1)=0$. Moreover, for $R/R_* > {\rm ln} \beta$ the scalar $f_R$ is strictly positive, and vanishes at $R/R_*= {\rm ln} \beta$. Thus, taking $0<\beta<1$ ensures that the minimum at $R_1=0$ of $V(R)$ never coincides with the value where $f_R$ vanishes. In fact, for $0<\beta<1$ the scalar $f_R= 1-\beta e^{-R/R_{*}}$ is positive definite in the domain $R\in [0,\infty)$, although $f_R$ could vanishes if $R$ becomes negative enough (see Figure~\ref{fig:f-f1}). However, in all the cases 
that we analyzed the cosmological dynamics is such that $R$ is never as negative as to fall into this pathology. Finally, we stress that $f_{RR}= \beta e^{-R/R_{*}}/R_*$ is positive definite for $\beta >0$.

In the past we have shown preliminary evidence that a viable cosmology is possible for this model~\cite{JaimeExp} when $\beta\geq 1$. Notably, when $\Lambda_{\rm eff}\neq 0$ ($\beta >1$). Nevertheless, as far as we are aware \cite{Exponential}, there are not detailed studies about the viability of 
$f(R)$ models allowing for asymptotically Ricci flat solutions\footnote{Strictly speaking the GR model with $f_{\rm GR}(R)= R-2\Lambda$ does not 
allow for asymptotically Ricci flat solutions.}. The goal of this paper is to fill that gap. In particular, we show that even if in this specific model $\Lambda_{\rm eff}\equiv 0$, the Ricci scalar $R$ has nevertheless a transient behavior which allows for an adequate accelerated expansion at present time, but with $R\rightarrow 0$ as $t\rightarrow \infty$. 
As we mentioned above, this is possible when $0<\beta\leq 1$. Here we report the capabilities of the Ricci flat scenarios in a FRW cosmology and in the construction of relativistic stars.

\begin{figure}
\begin{center}
\includegraphics[scale=0.7]{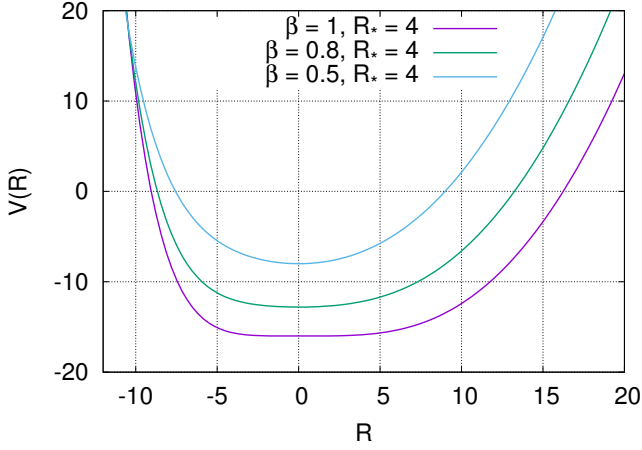}
\caption{The potential $V(R)$ (in units of $H_0^4$) associated with the exponential $f(R)$ model 
for $\beta = 0.5, \, 0.8$ and $1$ ($R$ in units of $H_0^2$). Notice that in each case the potential $V(R)$ has one critical point (a global minimum at $R=0$) 
corresponding to the value reached by $R$ at late cosmic times and also in asymptotically flat spacetimes. 
The minimum at $V(0)$ grows as $\beta\rightarrow 0$, reaching the maximum value  
$V(0)= 0$ in this limit. The minimum-minimorum is reached at $\beta=1$ when $\beta\in (0,1]$.}
\label{fig:Potentials}
\end{center}
\end{figure}

The potential $V(R)$ associated with this model is depicted in Figure~\ref{fig:Potentials} for three different values of $\beta \in (0,1]$. In this domain of $\beta$ the potential exhibit only a (global) minimum at $R=0$, which corresponds to the asymptotic value reached in cosmology at late times (see Section~\ref{sec:cosmologysols}), and also to the asymptotic value reached at spatial infinity (see Section~\ref{sec:SSSsols}). As $\beta\rightarrow 1$ the potential ``flattens'' around $R=0$, 
which makes the Ricci scalar $R$ to reach its asymptotic value $R=0$ monotonically and ``slowly'' in cosmic time. On the other hand, as $\beta$ decreases from $\beta=1$, the potential becomes like one of a harmonic oscillator near the minimum and $R$ can oscillate in cosmic time near $R=0$. 
These oscillations are otherwise damped due to the ``friction'' term generated by the expansion of the Universe (see Section~\ref{sec:cosmologysols}).

For $\beta>1$ the potential $V(R)$ develops several critical points, one being a global minimum which is associated with a de Sitter point~\cite{JaimeExp}.

In the following Sections we will analyze the cosmological scenario and the static and spherically symmetric spacetime, respectively.


\section{Cosmology}
\label{sec:cosmology}

We focus on a Friedmann-Robertson-Walker spacetime,

\begin{equation}
\label{SSmetric}
ds^2 = - dt^2  + a^2(t) \left[ \frac{dr^2}{1-k r^2} + r^2 \left(d\theta^2 + \sin^2\theta d\varphi^2\right)\right]\,\,\,, 
\end{equation}
where $k=\pm 1,0$, and for simplicity analyze only the case $k=0$. 

We also assume that the energy-momentum tensor of matter $T_{ab}$, described by a perfect-fluid, is a mixture of dark matter, baryons and radiation but in an epoch 
where the interaction between them can be neglected. 

Under these assumptions, Eqs. (\ref{traceR}) and (\ref{fieldeq3}) lead respectively to,
\begin{eqnarray}
\label{traceRt}
&& \ddot R = -3H \dot R - \frac{3f_{RRR} \dot R^2 + 2f- f_R R-\kappa (\rho_{\rm bar}+ \rho_{\rm DM}) }{3 f_{RR}} \nonumber \\
&& \,\,\,\\
\label{Hgen}
&& H^2 = \frac{\kappa}{3}\left(\rule{0mm}{0.3cm} \rho +\rho_{X}\right) \,\,\,,\\
\label{Hdotgen}
&& \dot{H}= -H^2 -\frac{\kappa}{6}\left\{\rule{0mm}{0.4cm} \rho +\rho_{X}+3\left(p_{\rm rad}+ p_{X}\right) \right\} \,\,\,,\\
\label{Hubble}
&& H = \dot a/a \,\,\,,
\end{eqnarray}
where $\dot{}\,\,= d/dt$ and 
\begin{eqnarray}
\label{rhoX}
\rho_X&=& \frac{   \frac{1}{2}(f_{R}R-f)-3f_{RR}H\dot{R} + \kappa\rho(1-f_{R}) }{\kappa f_{R}},\\
\label{pressX}
p_X &=& \frac{   \frac{1}{2}(f_{R}R-f)+3f_{RR}H\dot{R} - \kappa(\rho-3p_{{\rm rad}}f_{R}) }{3\kappa f_{R}},
\end{eqnarray}
are the density and pressure of the {\it geometric dark energy} (GDE), respectively. 

As usual, the matter variables obey their own dynamics provided by $\nabla_a T^{ab}_I=0$, where $I=1-3$, for each matter component 
(baryons, dark matter and photons) which for the actual case  leads to the standard conservation equation 
$\dot \rho_I - 3H (\rho_I + p_I)=0$. So $T_{ab}=\sum_{I=1}^{3} T^{ab}_I$. In the above equations $\rho = \sum_{I=1}^{3} \rho_{I}$.
The corresponding equations of state (EOS) are 
$p_{\rm bar,DM}=0$ for baryons and dark matter, $p_{\rm rad}=\rho_{\rm rad}/3$ for photons, and the $X$--fluid variables~(\ref{rhoX}) and~(\ref{pressX}), 
which satisfy a similar conservation equation, has the following EOS (for a thorough discussion about EOS of GDE in $f(R)$ see \cite{Jaime2014}):
\begin{equation}
\label{EOSX1b}
\omega_{X}= \frac{p_{X}}{\rho_{X}}=
\frac{3H^2-3\kappa\,p_{\rm rad}-R}{3\left(3H^2-\kappa\rho\right)}= \frac{1-\Omega_{\rm rad} -R/(3H^2)}{3 \Omega_X}\,\,\,,
\end{equation}
which evolves in cosmic time. Here $\Omega_X= \kappa \rho_X/(3 H^2)$ and $\Omega_{\rm rad}= \kappa \rho_{\rm rad}/(3 H^2)$. 
Similar fractional (dimensionless) energy-densities will be defined for the other matter components such that $\Omega_X+\sum_{I=1}^{3} \Omega_I=1$. 
Hereafter $\Omega_{\rm matt}:= \sum_{I=1}^{3} \Omega_I= \Omega_{\rm rad} + \Omega_{\rm bar}+ \Omega_{\rm DM}$. 
The {\it total} EOS is defined by 
 
\begin{eqnarray}
\label{EOSTOT}
\omega_{\rm tot} &:=& \frac{p_{\rm tot}}{\rho_{\rm tot}}= \frac{p_{\rm rad}+p_{X}}{\rho +\rho_{X}}= \nonumber \\
& &-\frac{1}{3}\left[ \frac{\frac{1}{2}\left(f_{R}R+f \right) + 3f_{RR}H\dot{R}-\kappa\rho}
          {\frac{1}{2}\left( f_{R}R-f\right) -3f_{RR}H\dot{R} + \kappa \rho}\right] \,\,\,,
\end{eqnarray}
where~(\ref{rhoX}) and~(\ref{pressX}) were used in the last equality. The total EOS is directly related with the deceleration parameter by 
($k=0$)
\begin{equation}
\label{EOSTOT2}
q := -\frac{\ddot a}{aH^2}= \frac{1}{2}\left(1+ 3 \omega_{\rm tot}\right) \,\,\,.
\end{equation}
From this equation we see that acceleration and deceleration occur when $\omega_{\rm tot}<-1/3$ and $\omega_{\rm tot}>-1/3$, respectively. 
In particular, when $\omega_{\rm tot}\rightarrow -1$, the Universe is dominated by dark-energy.

With all these ingredients we are able to analyze numerically the previous equations following the strategy detailed in~\cite{Jaime2012a}. 
This analysis in presented in the following Section~\ref{sec:cosmologysols}.


\subsection{Asymptotically Ricci flat solutions in cosmology}
\label{sec:cosmologysols}

As we discussed in the Introduction, one of the features that makes $f(R)$ gravity appealing is that it can produce an accelerated expansion at late times 
as generated purely by geometry due to the emergence of an effective cosmological constant $\Lambda_{\rm eff}= R_1/4$ from the dynamics of $R$. 

\begin{figure}
\begin{center}
\includegraphics[scale=0.65]{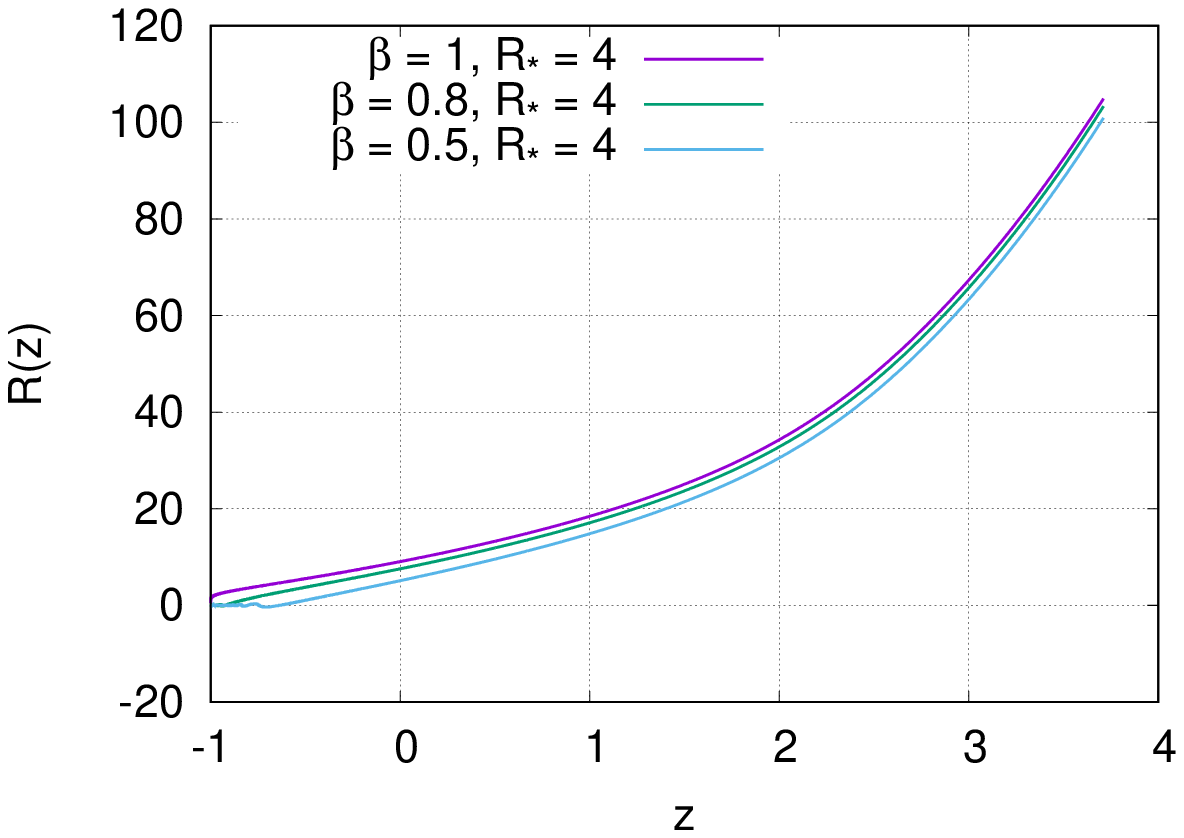}
\includegraphics[scale=0.65]{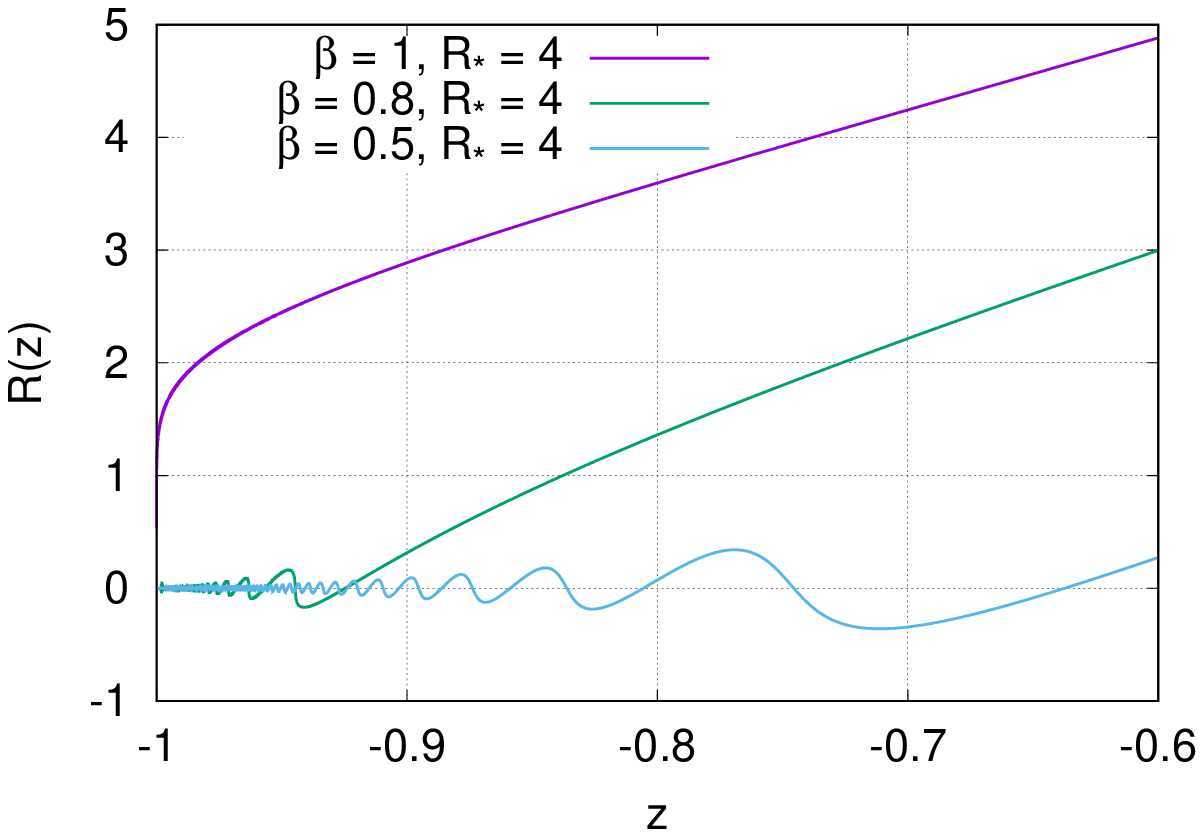}
\caption{Top: Ricci scalar (in units of $H_0^2$) versus ``red'' shift $z$ for $\beta=0.5, \, 0.8$ and $1$ with $R_*= 4H_{0}^{2}$.
Bottom: same as the top panel but for  $z\sim -1$ which is associated with the ``far future''.}
\label{fig:R-plano}
\end{center}
\end{figure}

\begin{figure}
\begin{center}
\includegraphics[scale=0.6]{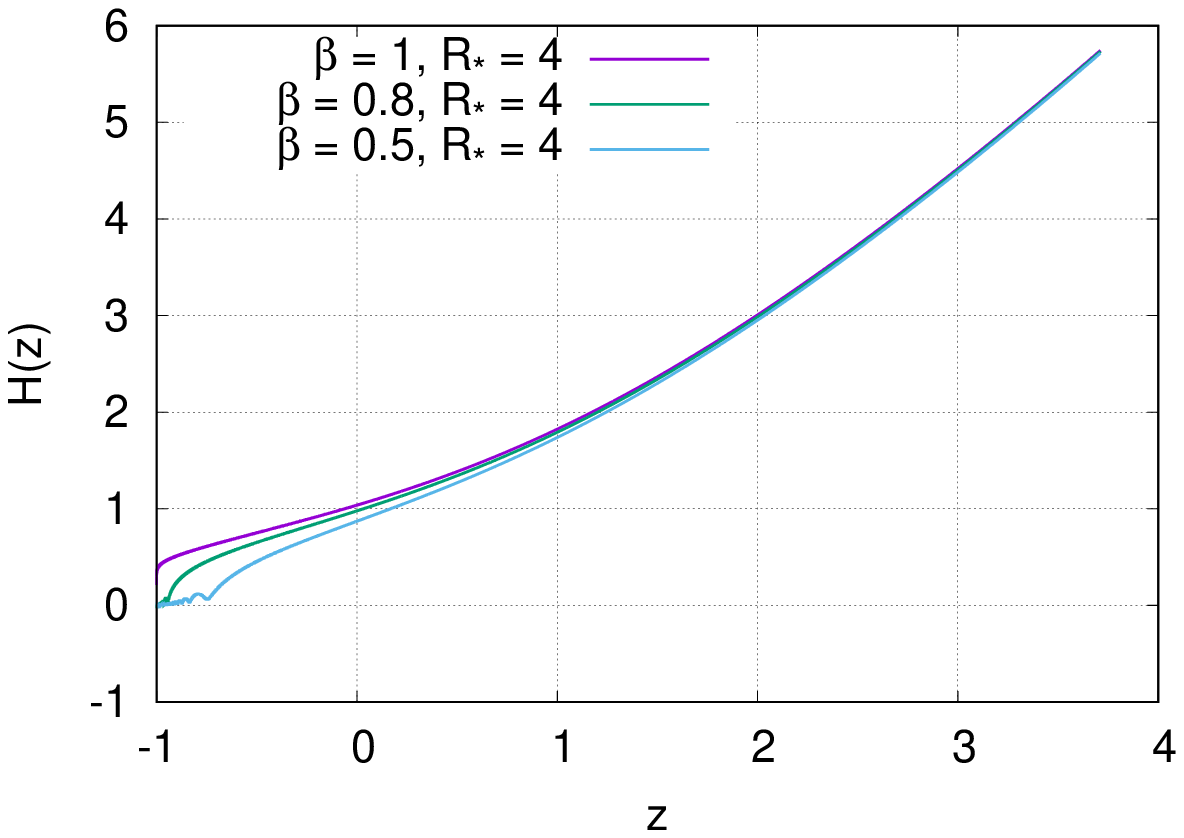}
\includegraphics[scale=0.6]{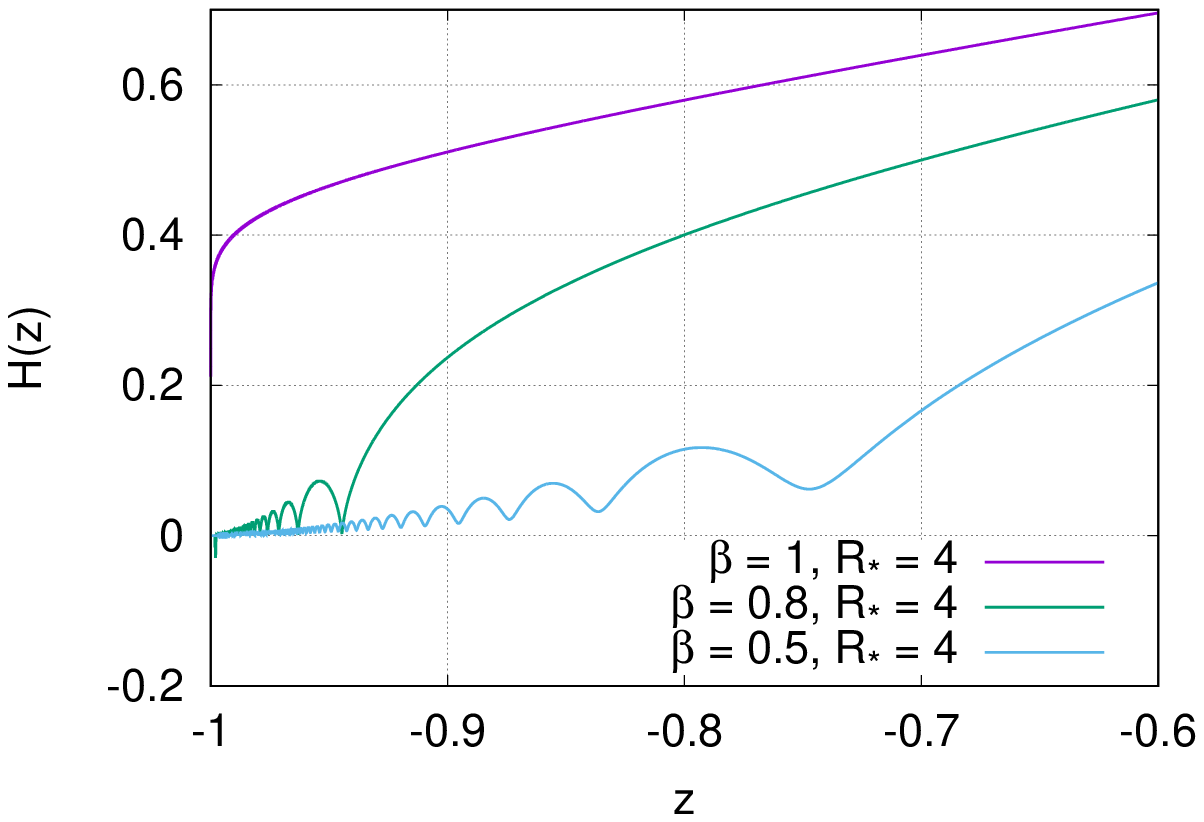}
\caption{Top: Hubble expansion (in units of $H_0$) versus $z$ associated with the cosmological solutions depicted in Figure~\ref{fig:R-plano}. 
Bottom: Hubble expansion for  $z\sim -1$. Notice that $H\rightarrow 0$ as $z\rightarrow -1$.}
\label{fig:H-plano}
\end{center}
\end{figure}

By integrating numerically the equations of Section~\ref{sec:cosmology} we show that it is possible to have a viable cosmology even when $\Lambda_{\rm eff}\equiv 0$. The expansion of the Universe under this particular dynamics can provide a matter dominated epoch followed by an accelerated one, even when asymptotically solutions goes to $R \rightarrow 0$. Figures ~\ref{fig:R-plano} and ~\ref{fig:H-plano} depict the Ricci scalar and the expansion rate $H$, respectively, during the cosmological evolution for two prototype values $\beta=1$ and $\beta=0.8$. Let us focus first on the value $\beta=1$. In this case, $R$ ``rolls'' slowly and monotonically towards the value $R=0$ through the potential $V(R)$ (cf. Figure~\ref{fig:Potentials}). When $R$ reaches the flat region of the potential a slowly varying effective cosmological ``constant'' appears and produces a transient accelerated expansion similar to the one observed at present time. Notice that the Hubble expansion, as depicted in Figure~\ref{fig:H-plano}, evolves from a high value at $z\gg 1$  (where $z=\frac{1}{a/a_0}-1$ and $a_{0}$ is the value of the scale factor today) to a vanishing value at very late times $z\rightarrow -1$, corresponding to $t\rightarrow \infty$. This means that the scale factor reaches a maximum when $t\rightarrow \infty$. This is reminiscent of the late-time behavior associated with a matter dominated model in GR (i.e. one without a cosmological constant and $k=0$ ) where $H^2~\sim a^{-3}= (z+1)^3$, which vanishes as $z\rightarrow -1$, i.e., $a\rightarrow \infty$. 

\begin{figure}
\begin{center}
\includegraphics[scale=0.6]{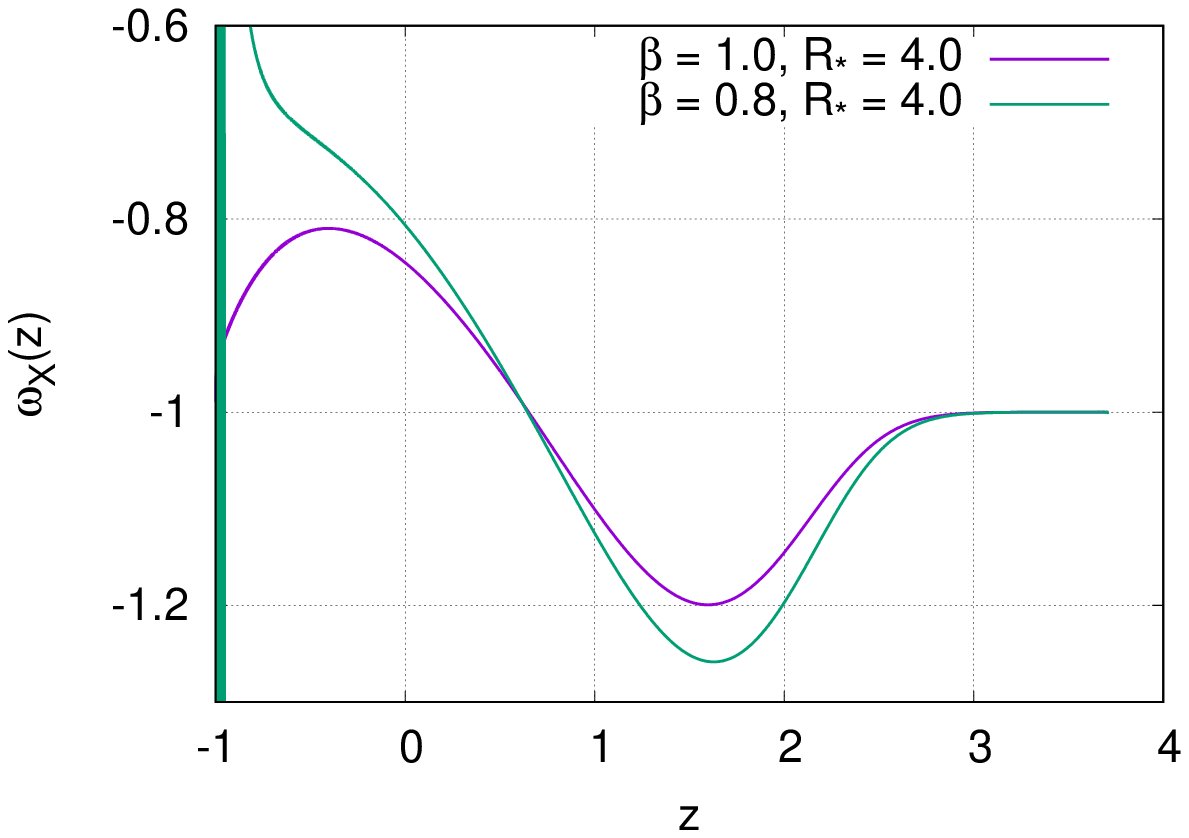}
\includegraphics[scale=0.6]{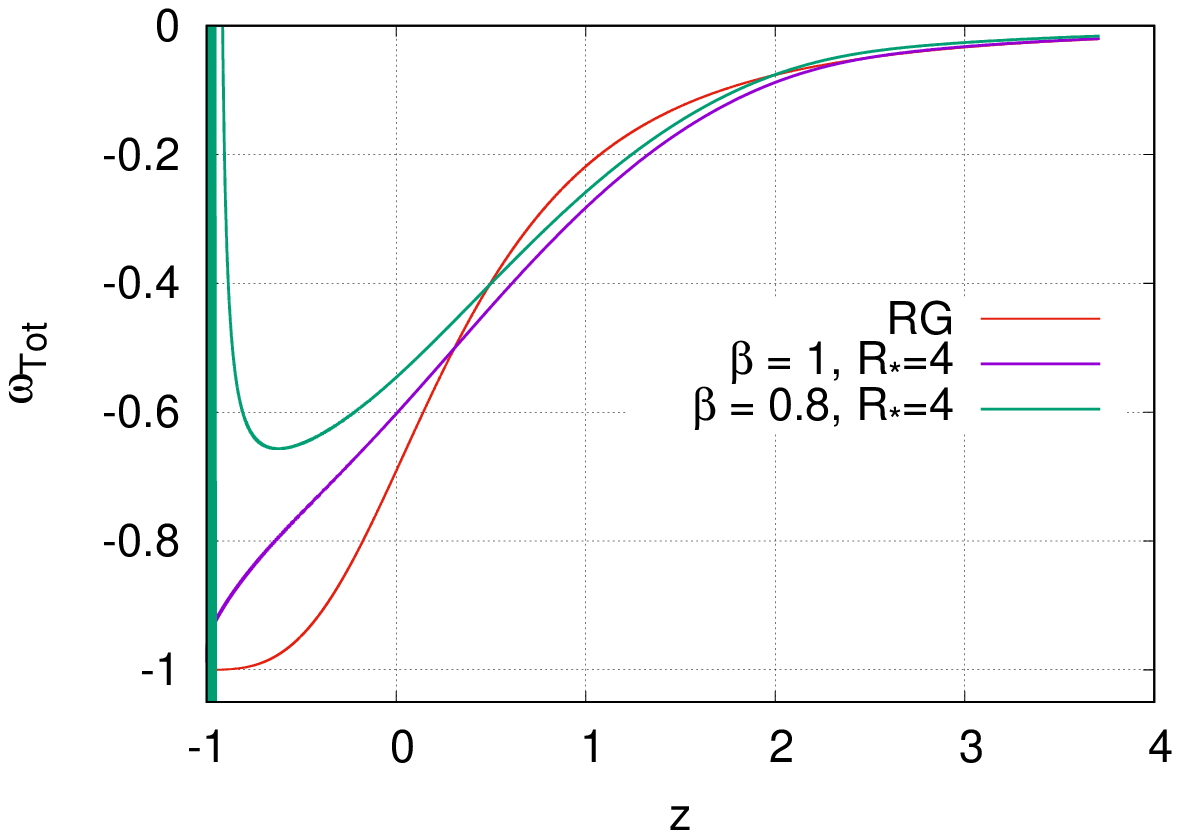}
\caption{Top: Evolution of the EOS associated with the geometric dark energy. Bottom: Evolution of the total EOS and their comparison with the 
$\Lambda$CDM evolution (labeled as GR).}
\label{fig:EOS}
\end{center}
\end{figure}

However, unlike the GR scenario, in this $f(R)$ model we have the following behavior: a radiation dominated epoch followed by a matter dominated era, both with a decelerated period $\ddot a<0$, associated with $\omega_{\rm tot}>-1/3$ (cf. Figure~\ref{fig:EOS}), then an accelerated period with $\ddot a>0$ corresponding to the epoch where the GDE dominates and which is associated with $\omega_{\rm tot}>-1/3$, and finally, a future period where again the 
Universe decelerates until the expansion stops at infinite time (cf. Figure~\ref{fig:EOS}).

The bottom-panel of Figure~\ref{fig:H-plano} zooms the behavior of $H$ at late times ($z\sim -1$). Notice that for $\beta = 1$ the value $f_R(R=0)=0$ is reached at $t\rightarrow \infty$.

\begin{figure}
\begin{center}
\includegraphics[scale=0.6]{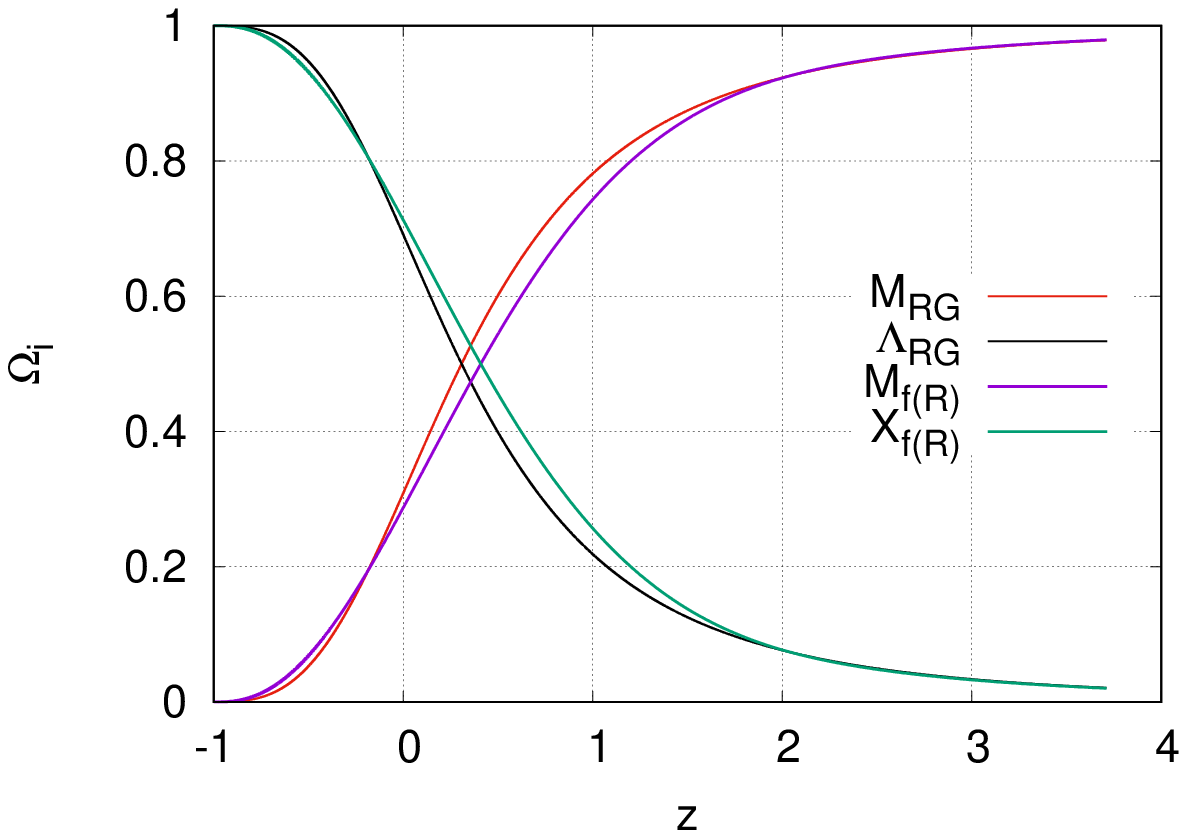}
\includegraphics[scale=0.6]{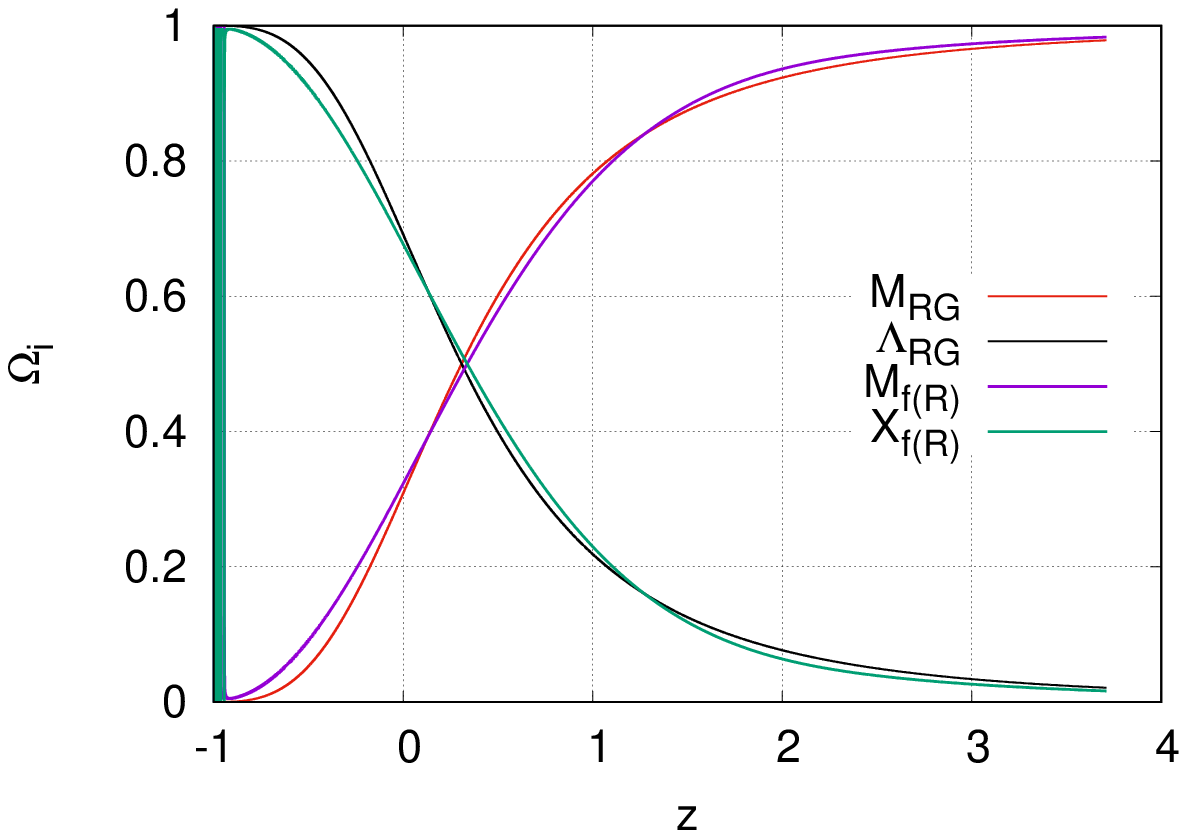}
\caption{Dimensionless density fractions $\Omega_X$ (geometric dark energy) and $\Omega_{\rm matt}$ (matter) 
for $\beta = 1$ (top panel) and $\beta = 0.8$ (bottom panel), taking $R_* = 4H_{0}^{2}$ and their comparison with the $\Lambda$CDM evolution (labeled as GR).}
\label{fig:Ome-todo}
\end{center}
\end{figure}

As concerns the value $\beta=0.8$, the dynamics change dramatically as compared to $\beta= 1$. The potential $V(R)$ behaves more like the potential of a harmonic oscillator. In this case $R$ ``rolls'' down the potential from a high value $R$ at $z\gg 1$ and pass beyond the minimum at $R_1=0$, despite the friction term, and start oscillating about the minimum. Notice however, that the oscillations start only in the future $z<0$, and prior to that, when $R>0$, an accelerated expansion takes place. During this period the  GDE contribution dominates over its matter counterpart (cf. Figure~\ref{fig:Ome-todo}).

Although $R$ can become negative during the oscillating period, the maximum negative amplitude of the oscillations is small enough to prevent a negative or zero $f_R$ where the equations become ill-defined. The amplitude is then damped due to the expansion of the Universe and ultimately $R$ vanishes asymptotically in time ($z\rightarrow -1$). 

The oscillating  behavior of $R$ is imprinted in $H$. The expansion $H$ oscillates in the future and the amplitude damps and vanishes as $z\rightarrow -1$. The bottom-panel of Figure~\ref{fig:H-plano} gives the impression that $H$ vanishes at the minima, but this is not the case. It vanishes only at infinite times.

Figure~\ref{fig:Ome-todo} depicts the dimensionless density fractions for matter and geometric dark energy, for $\beta=1$ (top panel) and $\beta=0.8$ (bottom panel). In both cases, we appreciate the matter dominated epoch $\Omega_X< \Omega_{\rm matt}$ and the GDE dominated era $\Omega_X> \Omega_{\rm matt}$. We remind the reader that the matter density $\Omega_{\rm matt}$ includes radiation, baryons and dark matter. This behavior is very similar to the $\Lambda$CDM model, in particular, notice that at present time ($z=0$) $\Omega_X\sim 0.7$ and $\Omega_{\rm matt}\sim 0.3$. 

\begin{figure}
\begin{center}
\includegraphics[scale=0.6]{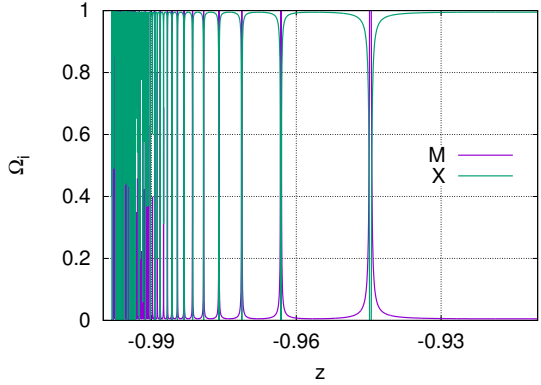}
\caption{Similar to the bottom panel of Figure~\ref{fig:Ome-todo} but for $z\sim -1$.}
\label{fig:Ome-zoom}
\end{center}
\end{figure}

Figure~\ref{fig:Ome-zoom} zooms the bottom panel of Figure~\ref{fig:Ome-todo} near $z=-1$. The oscillating behavior of the densities, induced by the oscillating behavior of $R$ and $H$, is clearly appreciated in this figure and also the way the matter and GDE dominates one over the other in alternating fashion.

The EOS for the GDE $\omega_X$ is shown in the top panel of Figure~\ref{fig:EOS} for $\beta = 1$ and $\beta = 0.8$. We appreciate that $\omega_X$ is not constant and is close to $-1$, notably, in past. As the Universe evolves, the EOS is below the so-called ``phantom divide'' region ($\omega_X<-1$) and then cross it to become $\omega_X\sim 0.8$ near the present time ($z=0$). This behavior of $\omega_X$ is characteristic of several $f(R)$ models~\cite{Jaime2014}, like the Hu--Sawicki~\cite{Hu2007} and the Starobinsky~\cite{Starobinsky2007} models. The main difference here emerges in the far future where $R\rightarrow 0$, while in the other $f(R)$ models $R\rightarrow R_1$, leading to a nonvanishing effective cosmological constant.

The total EOS $\omega_{\rm tot}$ is depicted in the bottom panel of Figure~\ref{fig:EOS}. This quantity is directly related with the accelerated or decelerated expansion [cf. Eq.~(\ref{EOSTOT2})]. As mentioned before, we see that during the evolution the Universe undergoes several phases of deceleration 
$\omega_{\rm tot}>-1/3$ and acceleration $\omega_{\rm tot}<-1/3$. Thus, unlike the $\Lambda$CDM model and other $f(R)$ models that generate an effective cosmological constant (including the exponential one with $\beta>1$ ~\cite{JaimeExp}) where the Universe accelerates indefinitely in the future, in the exponential models with $0<\beta\leq 1$ the Universe decelerates again in the future until the expansion ceases.

Our results shows that this model is able to explain the accelerated expansion and other cosmological phases adequately, like 
the $\Lambda$CDM model and some of the $f(R)$ models with $\Lambda_{\rm eff}\neq 0$~\cite{Jaime2012a}.

Before ending the cosmological analysis, a final comment is in order. It is worth stressing that not all the $f(R)$ models that admit an asymptotic Ricci flat solution can produce a successful cosmology. This in part is due to the crossing of a singularity associated with the zeros of the scalar $f_{RR}$ which interposes between the high curvature regime of the early Universe and the low or zero curvature domain of the late-time Universe. For instance, this can happen in the Starobinsky and Hu-Sawicki models~\cite{Starobinsky2007,Hu2007}. Other models, like the ``popular'' $f(R)= R^n$ model, can be free of those singularities while admitting an asymptotic Ricci flat solution, nevertheless, such model is simply unable to reproduce a cosmological evolution compatible with observations~\cite{Amendola2007a,Amendola2007b,Amendola2007c,Jaime2013}.


\section{Static and spherically symmetric (SSS) spacetimes}
\label{sec:Stars}

In this paper we also analyze the existence of relativistic objects within the exponential $f(R)$ model that admit an asymptotically Ricci flat background, 
in particular, asymptotically flat spacetimes\footnote{SSS spacetimes with a Ricci scalar behaving asymptotically as $R~\sim 1/r^2$  have a solid deficit angle, and so, they are not asymptotically Minkowski (cf.~\cite{Jaime2016}). Thus the fact that the Ricci scalar vanishes asymptotically does not guarantee asymptotic flatness.}.
 
In order to perform this analysis we assume a SSS spacetime described by the following metric:
\begin{equation}
ds^2 = - n(r) dt^2  + m(r) dr^2+ r^2 \left(d\theta^2 + \sin^2\theta d\varphi^2\right), 
\end{equation}
The final form of the equations for $n(r)$ and $m(r)$ are~\cite{Jaime2011}:

\begin{eqnarray}
\label{mprime}
m' &=& \frac{m}{r(2f_{R}+rR'f_{RR})} \Biggl{\{} 2f_{R}(1-m)-2mr^2 \kappa T^{t}_{\,\,t} 
 \nonumber \\
&&\!\!\!\!\!\!\!\!\!\!\!\!\!\!\!\!\! +\frac{mr^2}{3}(Rf_{R}+f+2\kappa T) 
+ \frac{rR'f_{RR}}{f_{R}}\Bigl{[}\frac{mr^2}{3}(2Rf_{R}-f+\kappa T) \nonumber \\
&&\!\!\!\!\!\!\!\!\!\!\!\!\!\!\!\!\!
-\kappa mr^2(T^{t}_{\,\,t}+T^{r}_{\,\,r})+2(1-m)f_{R}+2rR'f_{RR}\Bigr{]} \Biggr{\}} \; , 
\end{eqnarray}
\begin{eqnarray}
\label{nprime}
 n' &=& \frac{n}{r(2f_{R}+rR'f_{RR})} \Bigl{[} mr^2(f-Rf_{R}+2\kappa T^{r}_{\,\,r}) \nonumber \\
&& +2f_{R}(m-1)-4rR'f_{RR}  \Bigr{]} \; ,\\
\label{nbiprime}
n'' &=& \frac{2nm}{f_{R}} \Bigl{[}  \kappa T^{\theta}_{\,\,\theta}-\frac{1}{6}(Rf_{R}+f+2\kappa T)
+ \frac{R'}{rm}f_{RR}\Bigr{]} \nonumber \\
&& + \frac{n}{2r}\Bigl{[}2\left(\frac{m'}{m}-\frac{n'}{n}\right)+\frac{rn'}{n}\left(\frac{m'}{m}
+\frac{n'}{n}\right)\Bigr{]} \; .
\end{eqnarray}

For this spacetime the equation for the Ricci scalar reads,
\begin{eqnarray}
\label{traceRr}
R'' &=& \frac{1}{3f_{RR}}\Big{[}m(\kappa T+2f- Rf_{R}) - 3f_{RRR}R'^2\Big{]} \nonumber \\
&& +\left(\frac{m'}{2m}-\frac{n'}{2n}-\frac{2}{r}\right)R' \;.
\end{eqnarray}

The Ricci scalar computed directly from the metric also satisfies
\begin{eqnarray}
\label{Rr}
 R &=& \frac{1}{2r^2n^2m^2}\Bigl{[}4n^2m(m-1)+rnm'(4n+rn')  \nonumber \\
&& -2rnm(2n'+rn'')+r^2mn'^2\Bigr{]} \; .
 \end{eqnarray}

As concerns the matter sector, we consider a perfect fluid
\begin{equation}
\label{ecu:fluido-perfecto}
T_{ab}= (\rho + p)u_a u_b + g_{ab} p \; .
\end{equation}
where $p(r)$ and $\rho(r)$, are functions of the coordinate $r$ solely.

The behavior of this fluid will be described by the modified Tolman-Oppenheimer-Volkoff equation 
which arises from the conservation equation $\nabla^a T_{ab}= 0$. 
In fact, this equation reads exactly as in GR prior the substitution of the explicit form for $n'$:
\begin{equation}
\label{TOV}
p'= -(\rho + p) n'/2n \; .
\end{equation}
This equation completes our set of differential equations. As concerns the EOS, for simplicity we assume an {\it incompressible} fluid where the energy-density is given by a step function. So, the energy density $\rho$ is a nonzero constant $\rho_0$ within the star, but vanishes outside. In this way, Eq.~(\ref{TOV}) can be integrated without given any further EOS.
 
In the future we plan to analyze the inclusion of more realistic EOS where the energy-density is not kept constant (e.g.  polytropes). 

The numerical integration of the equations presented in this section is performed following the approach of \cite{Jaime2011}. The results will be presented in the next Section \ref{sec:SSSsols}.


\subsection{Asymptotically flat solutions for relativistic objects}
\label{sec:SSSsols}

The existence of relativistic compact objects, like neutron stars, are expected to be supported by any viable theory of gravity. 
As concerns, $f(R)$ gravity, some models intended to explain the cosmological observations 
seem to fail in this attempt. As shown by some authors using the scalar-tensor approach to $f(R)$ gravity, a {\it singularity} in the Ricci scalar was encountered at some spacetime region (see  \cite{Kobayashi2008,Kobayashi2009,Babichev2009,Upadhye2009} for this kind of results). However, some other models did not exhibit that kind of singularity \cite{Miranda2009}. In fact, the scalar-tensor approach can be well defined provided $f_R$ is a monotonic function of $R$. 
Otherwise, the resulting scalar-field potential is not single valued. Moreover, if the Einstein frame is used, this frame can become ill-defined if 
$f_R$ vanishes. The singularity found by some authors was related with some of those issues. So when $f_{RR}$ and/or $f_R$ are not positive definite, special care 
must be taken as to define the exact domains where the scalar-tensor approach is valid. In view of this, we believe that it is more 
advisable to remain in the original frame with its corresponding variables without introducing any other {\it fundamental} scalar than $R$ itself. 
This is precisely what we have done in Section~\ref{sec:Stars}, where no conformal transformation whatsoever 
was used to obtain the equations for SSS spacetime, nor any scalar $\Phi(R)= f_R$ was promoted as fundamental. This step would entail to 
invert the latter equation so as to obtain $R=R(\Phi)$, leading to $g(\Phi):= f(R(\Phi))$. 
As stressed before, this requires $f_R$ to be a monotonic function of $R$, which it is not always the case.

In the past we used the equations of Section~\ref{sec:Stars} to compute compact objects using two $f(R)$ models 
embedded in a de Sitter background~\cite{Jaime2011}, and showed that they were free of singularities.

We proceed now to follow the same approach used in~\cite{Jaime2011}, but for the exponential model and look for asymptotically flat solutions, 
given the fact that the potential $V(R)$ 
vanishes at its minimum for $\beta \in (0,1]$. In particular, we focus on those values of $\beta$ used to construct the cosmological models of previous 
sections which also avoids the singularity in the equations at $f_R(0)=0$, namely $\beta \in (0,1)$.
Moreover, we restrict to the simplest case of homogeneous (incompressible) density fluid. As described in~\cite{Jaime2011} we impose regularity conditions 
at the origin $r=0$ and the value $R|_{r=0}$ is used as a {\it shooting} parameter. One then look for an adequate value $R|_{r=0}$ such that 
$R\rightarrow 0$ at spatial infinity $r\rightarrow \infty$. Moreover, the Arnowitt-Deser-Misner (ADM) mass 
associated with the configuration most converge to a finite value if the 
spacetime is {\it genuinely} asymptotically flat (as opposed to a divergent value if $R\sim 1/r^2$ asymptotically, as it usually happens in spacetimes having a solid deficit angle). 
We then solve numerically the differential equations (\ref{mprime})--(\ref{traceRr}) and (\ref{TOV}) to find $m(r)$, $n(r)$, 
$R(r)$ and $p(r)$, respectively, for $r\in [0,\infty)$. 
In principle, its is not necessarily to solve the second order Eq.~(\ref{nbiprime}), however, we also solve it and together with (\ref{Rr}) we check the self-consistency 
of our numerical results. Any bug or mistake in the numerical code would reflect in a lack of self-consistency in our solutions. 
This self-consistency is achieved within the 
accuracy of the 4th order Runge-Kutta (double-precision) 
FORTRAN algorithm that we employed to solve the system of equations. So given a sufficiently small integration step, 
the numerical consistency is found within the numerical errors associated with this algorithm 
${\cal O}(10^{-10})$. Furthermore, replacing (\ref{mprime})--(\ref{nbiprime}) in (\ref{traceRr}) leads to an identity $R\equiv R$. 
However, this identity cannot be taken for granted if a mistake is committed somewhere in the numerical code, notably, when introducing the 
differential equations in the FORTRAN language.

\begin{figure}
\begin{center}
\includegraphics[scale=0.6]{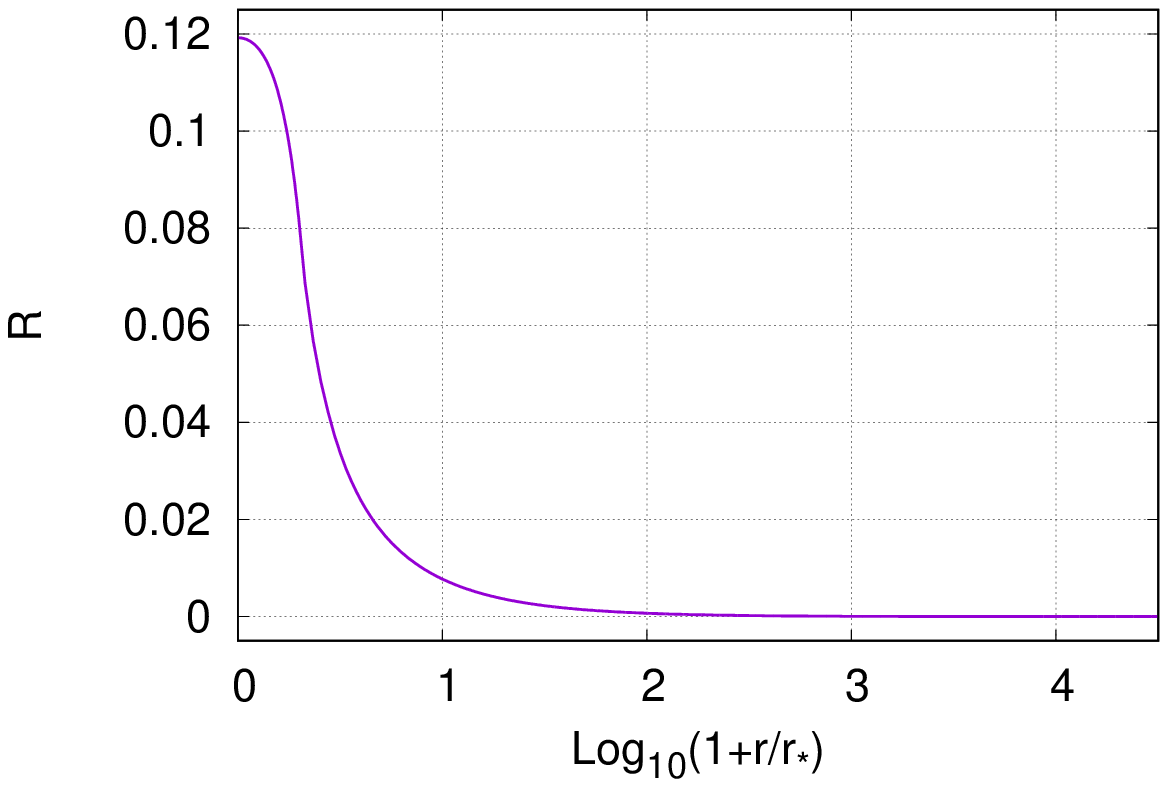}
\includegraphics[scale=0.6]{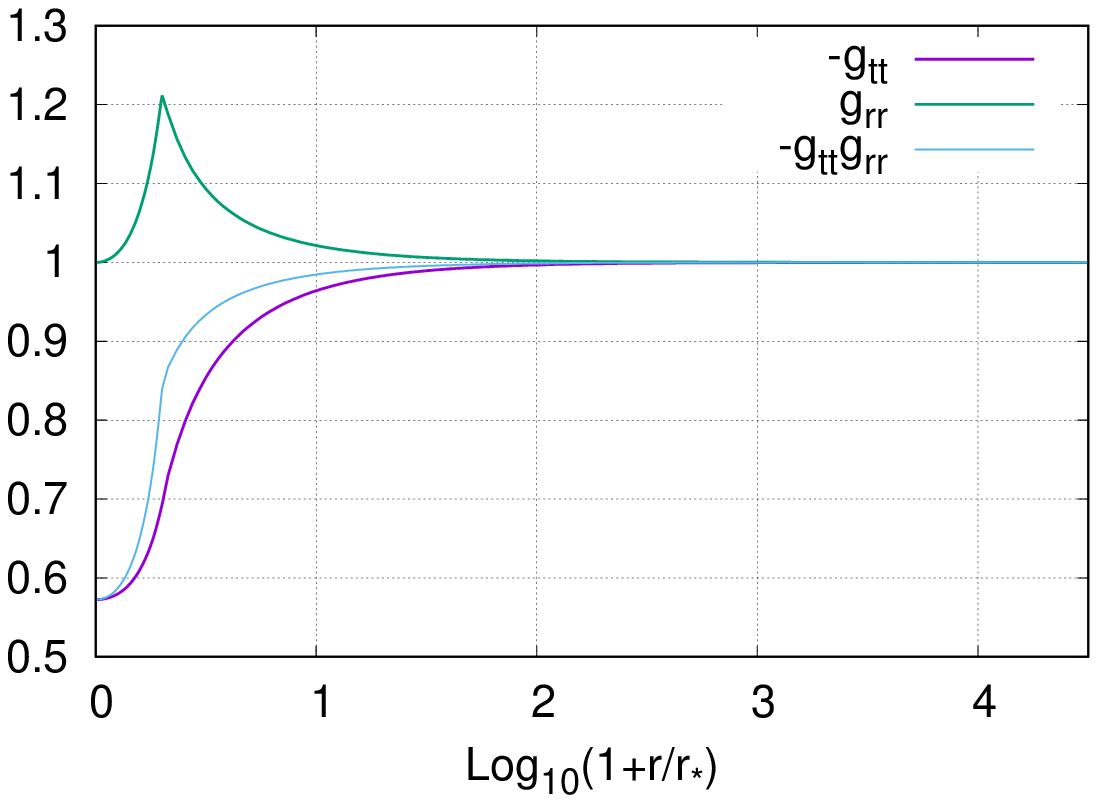}
\includegraphics[scale=0.6]{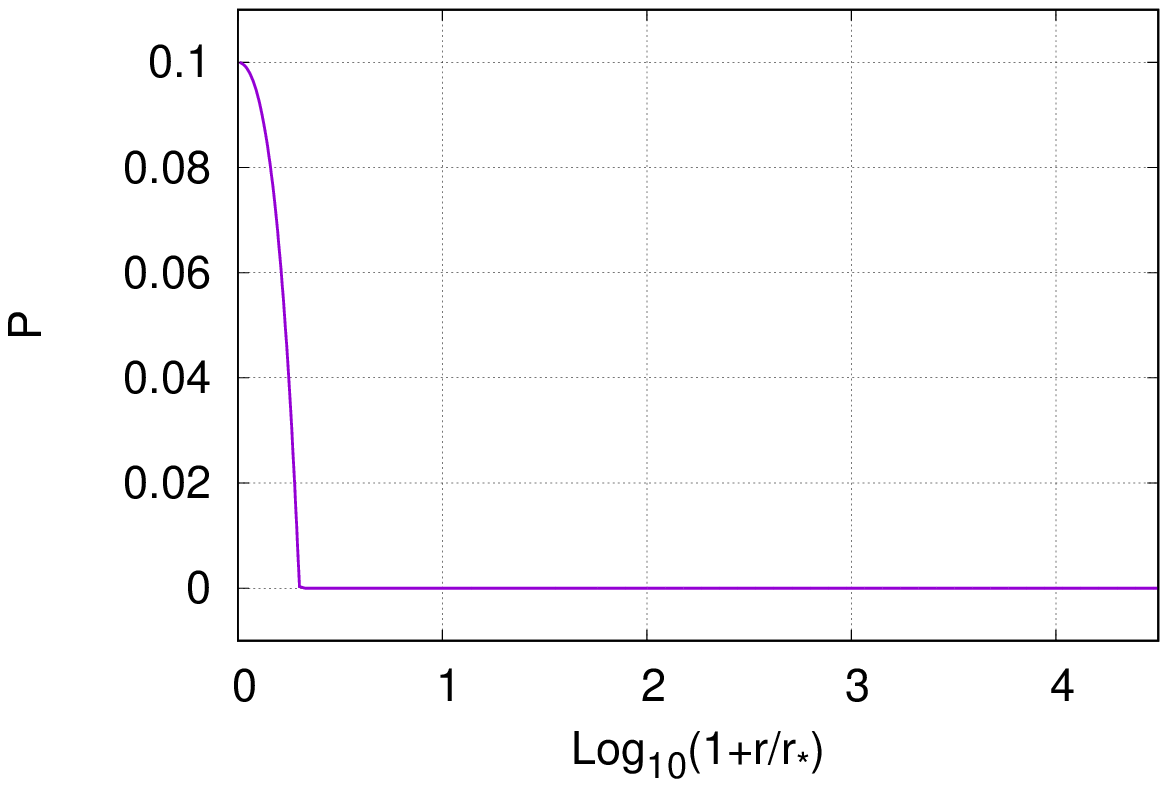}
\caption{Top: Ricci scalar (in units of $R_*$) as a function of $r/r_*$ 
for $\beta=0.8$. Here $r_*\sim 4.8 \times 10^{-4}$ is the (area) radius of the ``star'' (in units of $R_*^{-1/2}$ with 
$R_*= 4H_0^2$) defined to be where the pressure $p$ vanishes. Middle: the metric components $n=-g_{tt}$, 
$m=g_{rr}$ and $nm= -g_{tt} g_{rr}$.  Notice from these two panels the asymptotically-flat behavior of the spacetime. 
Bottom: fluid's pressure $p$ (in units of the constant density $\rho_0$) where $\rho=\rho_0=2.8 \times 10^{4} R_*/G_0$ 
for $r\in [0,r_*]$ and zero outside the object.}
\label{fig:stars1}
\end{center}
\end{figure}

The numerical results using this methodology are depicted in Figure~\ref{fig:stars1}. The Ricci scalar is always positive and interpolates 
monotonically between the center of the star at $r=0$ to spatial infinity without encountering any singularity. Thus, $f_R$ never vanishes in this spacetime, which precludes the equations to become singular when $f_R=0$. We remind the reader that $f_R$ only vanishes when $R<0$ for $\beta\in (0,1)$. The metric components have the typical form of a SSS spacetime generated by a globally regular compact object (see the middle panel of Figure~\ref{fig:stars1}). 
The fact that $n=-g_{tt}$ differs from unity at the center of the object ($r=0$) by several percent indicates that the gravitational field is strong there as $n$ represents the square of the so called {\it redshift} factor. 
The {\it cusp} produced in the metric component $m=g_{rr}$ at the star's surface $r_*$ 
where the pressure vanishes (see the bottom panel of Figure~\ref{fig:stars1})
is due to the discontinuity associated with the use of a step function for $\rho$. 
Nevertheless, this cusp can be smoothed out 
when using a more realistic EOS (e.g. a polytrope) which allows for a density to vary smoothly with $r$, like the pressure itself. 

The middle panel of Figure~\ref{fig:stars1} also depicts the product $-g_{rr} g_{tt}$. In GR 
this quantity is usually unity outside the star, where the Birkhoff theorem applies, and where the metric is given by the (vacuum) 
Schwarzschild solution. Thus, this product allows to appreciate the deviations of the metric from the Schwarzschild solution outside the object. These 
deviations are due to a nontrivial solution of the 
Ricci scalar outside the star. In GR the Ricci scalar is given by $R=-\kappa T= \kappa(\rho-3p)$. So for the constant-density model, 
$R$ grows monotonically from its central value $R=\kappa(\rho-3p)|_{r=0}< \rho$ to its value $R\approx \kappa \rho$ near the surface of the star where $p\ll\rho$, 
and then drops to zero outside the star in a discontinuous way. In the exponential model, $R$ varies smoothly from $r=0$ to spatial infinity where it vanishes. 
At spatial infinity $R=0$ is clearly a solution of Eq.~(\ref{traceRr}). 

\begin{figure}
\begin{center}
\includegraphics[scale=0.6]{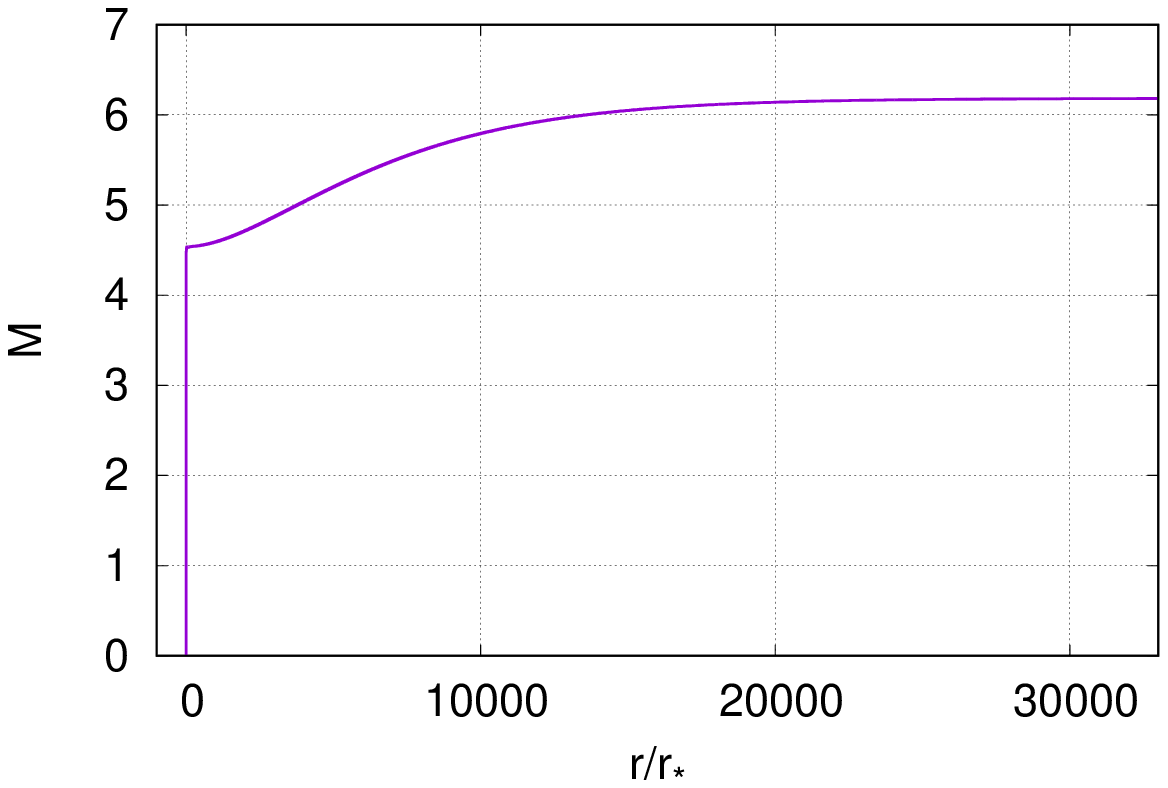}
\includegraphics[scale=0.6]{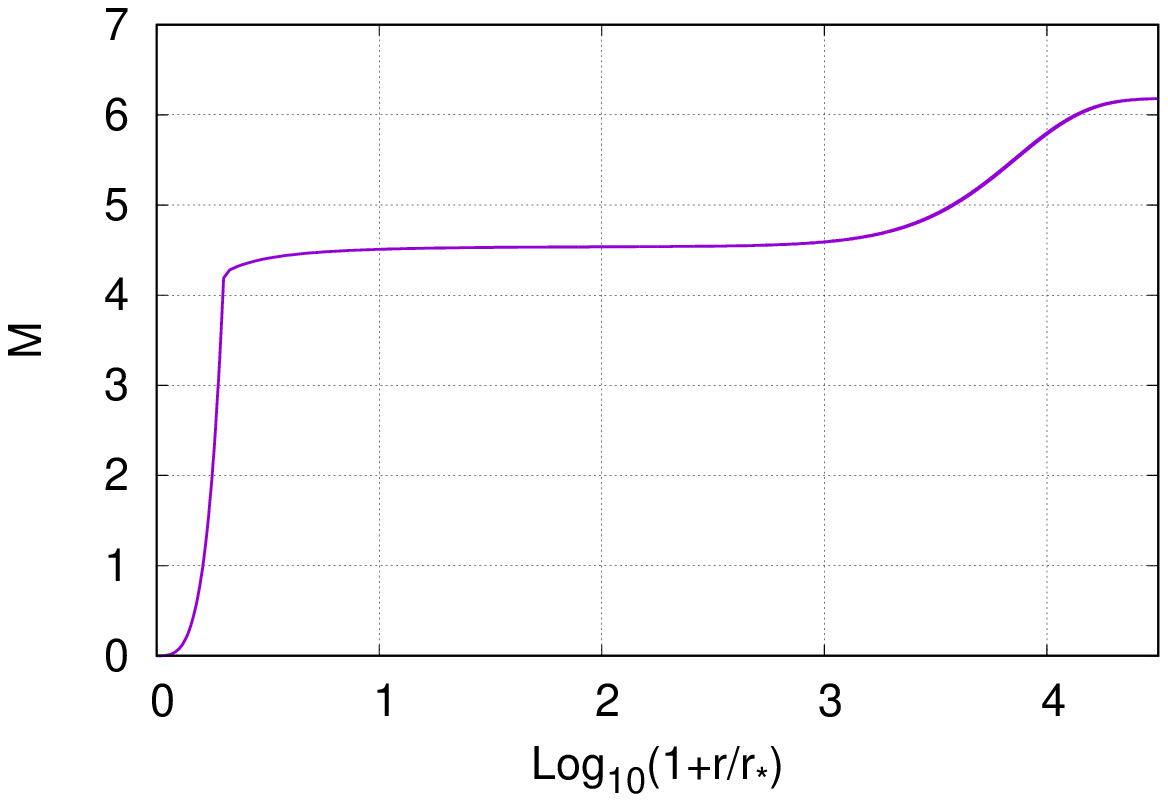}
\includegraphics[scale=0.6]{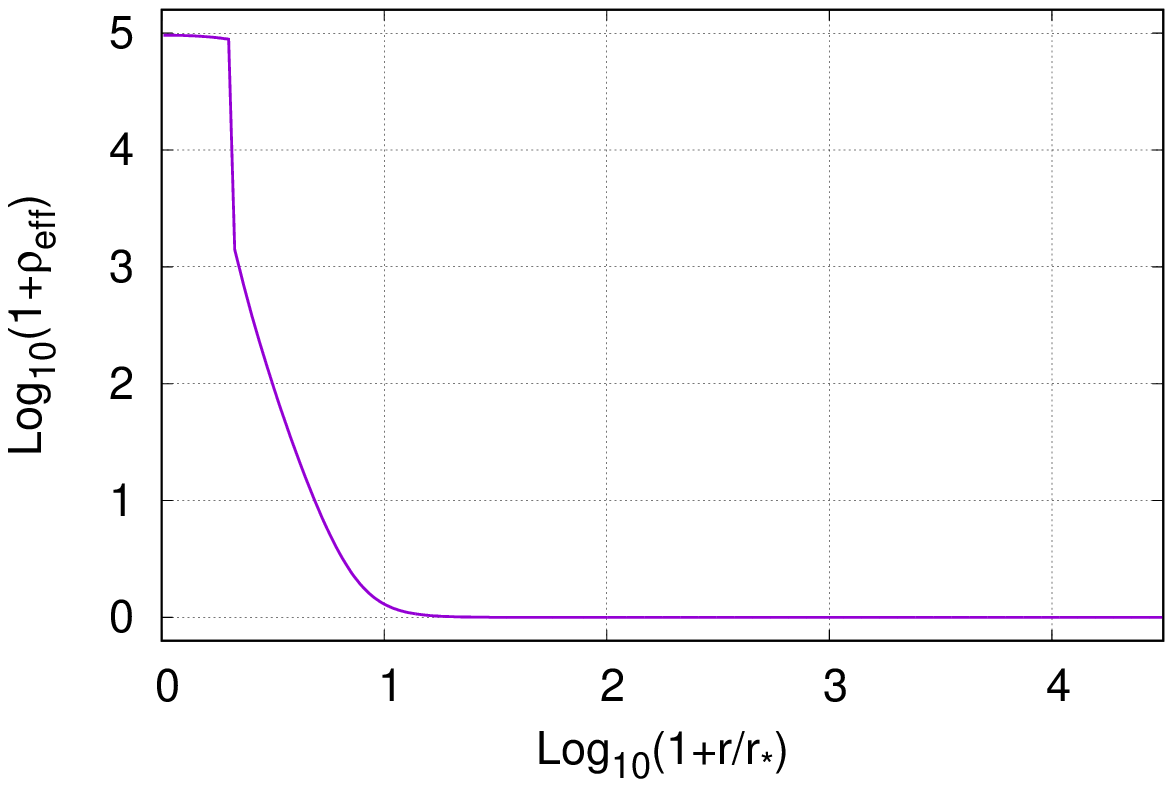}
\caption{Top: Mass function $M$ (in units of $10^{-5}c^2/(G_0R_*^{1/2})$) associated with the configuration of Figure~\ref{fig:stars1}. 
Middle: similar to the top panel but using a logarithmic variable on the horizontal axis in order to zoom out the inner region of the star.
The ADM mass ${\cal M}$ corresponds to the value $M$ at $r\rightarrow \infty$. Bottom: effective energy-density (in units 
of $R_*/G_0$) as defined in the main text. Unlike the pressure $p$ and the density $\rho$ of the fluid, 
this density does not vanish exactly for $r>r_*$, i.e., 
outside the compact support of the star $r\in [0,r_*]$; it vanishes asymptotically like 
the Ricci scalar (cf. the top panel of Figure~\ref{fig:stars1}).}
\label{fig:Madm}
\end{center}
\end{figure}

The $g_{rr}=m$ component can be used to compute the ADM mass 
${\cal M}$ of the object from the parametrization $g_{rr}(r)= [1 -2M(r)G_0/r]^{-1}$. 
That is, $M(r)=r(m-1)/(2G_0m).$ So ${\cal M}=M(\infty)$. Notice from Figure~\ref{fig:Madm} (top and middle panels) 
that the mass function $M(r)$ converges to the (ADM) mass ${\cal M}$ of the configuration as $r\rightarrow \infty$, 
and unlike the GR scenario, $M(r)$ grows outside the compact support of the star ($r>r_*$) due to 
the contributions associated with the effective energy-density $\rho_{eff}$ which extend outside the star.
Notably, by the contributions of several $f(R)$ quantities which extend beyond $r_*$. 
The density $\rho_{eff}$, which includes the fluid's density $\rho$, 
can be obtained from the (total) effective 
energy-momentum tensor that one can define within $f(R)$ gravity [cf. the right-hand-side 
of Eq.~(\ref{fieldeq3}) and~Ref.~\cite{Jaime2014} for a discusion], or more explicitly, from  
Eq.~(\ref{mprime}). That is, from Eq.~(\ref{mprime}) one can write an equation for $M(r)$ in the form $M'= 4\pi \rho_{eff} r^2$ 
from which $\rho_{eff}$ can be readoff. 
Figure~\ref{fig:Madm} (bottom panel) depicts $\rho_{eff}(r)$ showing that it vanishes asymptotically like the Ricci scalar 
(cf. the top panel of Figure~\ref{fig:stars1}). Remarkably, this density turns out to be nonnegative which explains the 
monotonically grow of $M(r)$.

An open problem that remains to be investigated is the study of the stability of 
stars in the framework of $f(R)$ gravity. This issue requires a separate analysis and is currently under scrutiny~\cite{AlcubierreNosotros}.
\bigskip

\section{Discussion}
\label{sec:discussion}
In this paper we analyze the viability of a FRW cosmology within the framework of an exponential $f(R)$ model where the effective cosmological constant vanishes asymptotically in cosmic time. The accelerated expansion is produced by a transient behavior of the Ricci scalar that allows for a sufficiently long (geometric) dark energy domination that follows after the matter dominated era. Depending on the parameters of the model, this dark-energy domination drops in the far future in a monotonic way or in an oscillating fashion until the Universe stop expanding. All these features can be summarized by looking to the total EOS, which behaves in different ways according to the dominating type of component (e.g. matter or dark energy). The resulting cosmology does not differ significantly from the $\Lambda$CDM model of GR. However, the exponential model predicts a very specific variation on the effective EOS of dark energy, which can be confronted with future observations \cite{surveys}.

The current analysis was limited in several aspects. For instance, we did not attempt to best-fit the parameters of the model and the initial conditions used 
in the numerical integration by using actual data (like SNIa).  We did not confront the exponential model 
to cosmological perturbations, and thus, we were not able to study the anisotropies of the CMB and many other aspects associated with them. 
We plan to overcome these limitations in a future and more detailed work.

Finally, we explored the capability of the exponential model in the construction of relativistic compact objects within an asymptotically flat background 
by keeping the same values of the parameters used in the cosmological models with $\beta \in (0,1)$. We showed using a numerical analysis that such objects can indeed be constructed 
without finding any kind of singularity within a static and spherically symmetric spacetime. This analysis was also limited in several aspects, but 
allowed us to pave the way for a more delicate study using the same tools. For instance, on the strong gravity aspect, we expect to implement the use of more 
realistic equations of state for the nuclear matter and try to build actual neutron-star sized objects. On the weak gravity side, we plan to investigate if some sort of chameleon mechanism appears and allows for the exponential model to actually pass the Solar System tests. 
This analysis requires the handling of a huge numerical precision ~\cite{Hu2007} which is beyond the capabilities of the standard ``crunch-number'' 
programming.


\section*{Acknowledgments}

This work was supported in part by DGAPA--UNAM grants IN107113, RR107015 and SEP--CONACYT grants CB--166656 and CB--239639. 




\begin{thebibliography}{99}


\bibitem{Sotiriou2010}
T. P. Sotiriou and V. Faraoni, \Journal{\RMP}{82}{451}{2010}.

\bibitem{Capozziello2008a}
S. Capozziello, and M. Francaviglia, 
\Journal{\GRG}{40}{357}{2008}.

\bibitem{deFelice2010}
A. De Felice, and S. Tsujikawa, 
\Journal{\LRR}{13}{3}{2010}.

\bibitem{Jaime2012a}
L. G. Jaime, L. Pati\~no, and M. Salgado, arXiv: 1211.0015.

\bibitem{Starobinsky1979}
A. A. Starobinsky, JETP Lett. {\bf 30}, 682 (2007).

\bibitem{Perlmutter1999}
S. Perlmutter {\it et al.}, \Journal{\APJ}{517}{565}{1999}. 

\bibitem{Riess1998}
A. G. Riess {\it et al.}, \Journal{\AJ}{116}{1038}{1998}. 

\bibitem{Amanullah2010}
R. Amanullah {\it et al.}, (Supernova Cosmology Project), \Journal{\APJ}{716}{712}{2010}.

\bibitem{Jaime2016}
P. Ca\~nate, L. G. Jaime, and M. Salgado, \Journal{\CQG}{33}{155005}{2016}.

\bibitem{Jaime2014}
L. G. Jaime, L. Pati\~no and M. Salgado, \Journal{\PRD}{89}{084010}{2014}.

\bibitem{Faulkner2007}
T. Faulkner, M. Tegmark, E. F. Bunn, and Y. Mao, \Journal{\PRD}{76}{063505}{2007}.

\bibitem{Khoury2004}
L. Khoury and A. Weltman, \Journal{\PRL}{93}{171104}{2004}; {\it ibid}, \Journal{\PRD}{69}{044026}{2004}

\bibitem{Hu2007}
W. Hu, and I. Sawicky, \Journal{\PRD}{76}{064004}{2007}.


\bibitem{Kobayashi2008}
T. Kobayashi, and K. Maeda, 
\Journal{\PRD}{78}{064019}{2008}.

\bibitem{Kobayashi2009}
T. Kobayashi, and K. Maeda, 
\Journal{\PRD}{79}{024009}{2009}.

\bibitem{Babichev2009}
E. Babichev, and D. Langlois, 
\Journal{\PRD}{80}{121501(R)}{2009}; {\it idem}, arXiv: gr-qc/0911.1297

\bibitem{Upadhye2009}
A. Upadhye, and W. Hu,
\Journal{\PRD}{80}{064002}{2009}.

\bibitem{Jaime2011}
L. G. Jaime, L. Pati\~no, and M. Salgado,
\Journal{\PRD}{83}{024039}{2011}.

\bibitem{neutronstarsf(R)}
S. Yazadjiev, D. Doneva, K. Kokkotas, K. V. Staykov, \Journal{\JCAP}{06}{03}{2014}; L. Sagunski {\it et al.}, arXiv: gr-qc/1709.06634

\bibitem{boss2014}
T. Delubac {\it et al.}, \Journal{\AA}{574}{A59}{2015}.

\bibitem{Zhao2017}
G. B. Zhao {\it et al.}, \Journal{Nature Astronomy}{1}{627}{2017}.

\bibitem{Jaime2015}
L. G. Jaime, \Journal{\PRD}{91}{124070}{2015}.

\bibitem{Amendola2007a} 
L. Amendola, D. Polarski, and S. Tsujikawa, 
\Journal{\PRL}{98}{131302}{2007}.

\bibitem{Amendola2007b} 
L. Amendola, R. Gannouji, D. Polarski, and S. Tsujikawa, 
\Journal{\PRD}{75}{083504}{2007}.

\bibitem{Amendola2007c} 
L. Amendola, D. Polarski, and S. Tsujikawa, 
\Journal{\IJMPD}{10}{1555}{2007}.

\bibitem{Jaime2013} 
L. G. Jaime, L. Pati\~no, and M. Salgado, \Journal{\PRD}{87}{024029}{2013}.

\bibitem{Starobinsky2007}
A. A. Starobinsky, JETP Lett. {\bf 86}, 157 (2007).

\bibitem{Miranda2009}
V. Miranda, S. Jor\'{a}s, I. Waga and M. Quartin, \Journal{\PRL}{102}{221101}{2009}.

\bibitem{JaimeExp}
L. G. Jaime, M. Salgado and L. Pati\~no, \Journal{Springer Proc.Phys.}{157}{363}{2014}.


\bibitem{Exponential}
R. Kerner, \Journal{\GRG}{14}{453}{1982}; 
E. Elizalde, S. Nojiri, S. D. Odintsov, and S. Zerbini, \Journal{\PRD}{77}{046009}{2008}; 
L. Yang, C. C. Lee, L. W. Luo, and C. Q. Geng, \Journal{\PRD}{82}{103515}{2010};
K. Bamba, C. Q. Geng, and C. C. Lee, \Journal{\JCAP}{08}{021}{2010}; 
E. Elizalde, S. Nojiri, S. D. Odintsov, and S. Zerbini, \Journal{\PRD}{83}{086006}{2011}; 
E. Elizalde, S. D. Odintsov, L. Sebastiani, and S. Zerbini, arXiv: 1108.6184;
E. V. Linder, \Journal{\PRD}{80}{123528}{2009}.

\bibitem{AlcubierreNosotros}
J. C. Degollado, L. G. Jaime,  S. Joras, M. Salgado and M. Alcubierre 
(in preparation).

\bibitem{surveys}

R.~Laureijs {\it et al.} [EUCLID Collaboration]  
(ESA/SRE) arXiv: 1110.3193.  

L.~Amendola {\it et al.} [Euclid Theory Working Group Collaboration],  
arXiv: 1606.00180  

A. Aghamousa {\it et al.} [DESI Collaboration],  
(FERMILAB-PUB-16-517-AE) arXiv: 1611.00036.  


 



\end{thebibliography}
\end{document}